\documentclass[11pt, A4]{article}

\usepackage{graphicx}
\usepackage{amssymb}
\usepackage{amsthm}
\usepackage{subfigure}
\usepackage{url}
\usepackage{amsmath}
\usepackage[russian,english]{babel}
\usepackage{authblk}

\usepackage{anysize}
\marginsize{2cm}{2cm}{.3cm}{2cm}

\usepackage{fancyhdr}
\graphicspath{{figs/}}

\begin{document}

%%%% Article title to be placed here
\title{The dry history of liquid computers}

\author{Andrew Adamatzky}

\date{
Unconventional Computing Lab, \\
University of the West of England, Bristol, UK\\
\texttt{andrew.adamatzky@uwe.ac.uk}}

%%%%%%%%%%%%%%%%%%%%%%%%%%%

\maketitle

\begin{abstract}
A liquid can be used to represent signals, actuate mechanical computing devices and to modify signals via chemical reactions. We give a brief overview of liquid based computing devices developed over hundreds of years. These include hydraulic calculators, fluidic computers, micro-fluidic devices, droplets, liquid marbles and reaction-diffusion chemical computers.
\end{abstract}

%\tableofcontents

\section{Introduction}

A substance offering no resistance to a shear deformation is a fluid. A liquid is an incompressible fluid.  Our personal interest in liquid based computers started in early 1990s where we proposed a paradigm and experimental laboratory implementations of reaction-diffusion chemical computers~\cite{DBLP:conf/parcella/Adamatzky96,adamatzky2005reaction}, affective liquids~\cite{adamatzky2002dynamics} and liquid brains for robots~\cite{adamatzky2003liquid,adamatzky1999wet}. In reaction-diffusion computers information is processed via interaction of phase or diffusion wave fronts. Affective liquids are mixtures of chemical species representing emotional states~\cite{adamatzky2002dynamics}, and stirred or thin-layer mixtures of doxastic and affective chemical-like species~\cite{adamatzky2005parachemistry}. Liquid brain for robots are onboard controllers for robot navigation which employ a thin-layer excitable chemical reaction~\cite{adamatzky2005non}.

\begin{table}[!bp]
  \caption{A brief history of liquid computers}
    \centering
    \begin{tabular}{p{1cm}p{8cm}p{5cm}}
    Year    & Device & Publications \\  \hline
    1900    &  Hydraulic algebraic machines   &  \cite{emch1901two,gibb1914,frame1945machines}  \\
    1920   &   Tesla diode  & \cite{tesla1920valvular}\\
    1936    & Hydraulic integrators     &  \cite{moore1936hydrocal,luk1939hydraulic}  \\
     1949   & Monetary National Income Analogue Computer     &  \cite{bissell2007historical}  \\ 
     1949   & Fluid mappers & \cite{moore1949fields}    \\
     1960    & Fluidic logic  & \cite{hobbs1963fluid,peter1965and,paul1969fluid,drake1965pure,bowles1965passive}\\
    1985 & Belousov-Zhabotinsky computers &
    \cite{kuhnert1989image,agladze1996chemical,steinbock1996chemical, gorecki2003chemical,adamatzky2005reaction} \\
    1996 & Reaction-diffusion computers & \cite{adamatzky1996reaction,tolmachiev1996chemical,adamatzky1997chemical,adamatzky1994constructing}\\
    2003 & Liquid brain for robots & \cite{adamatzky2003liquid,adamatzky2004experimental}\\
    2003 & Fluid maze solver & \cite{fuerstman2003solving}\\
    2007 & Droplet  logic (pressure/flow driven) & \cite{cheow2007digital,fair2007digital,toepke2007microfluidic,prakash2007microfluidic}\\
2010 & Chemotactic droplets solving mazes & \cite{lagzi2010maze,cejkova2014dynamics} \\
2012 & Droplet logic  &   \cite{mertaniemi2012rebounding} \\
2017 & Liquid marbles logic  & \cite{draper2017liquid}
    \end{tabular}
    \label{tab:history}
\end{table}

A chronological order of selected liquid based computing devices is shown in Tab~\ref{tab:history}. Most of the prototypes are discussed in the paper.
In these devices liquid is used in various roles. Mass transfer analogies are employed in hydraulic mathematical machines in Sect.~\ref{hydraulicmachines} and integrators Sect.~\ref{hydraulicintegrations}. Flows of fluid explore and map templates, and  solve mazes, in fluid mappers, Sect.~\ref{fluidmappersmazesolvers}. In Sect~\ref{fluidmappersmazesolvers} the fluid flow is driven by the fluid pressure. A fluid flow can be also evoked by temperature and chemical concentration gradients, and then visualised by travelling liquid droplets, as shown in  Sect.~\ref{mazesolvingdroplets}. Interacting fluid jets realise logical gates in fluidic logic devices in Sect.~\ref{fluidlogic}, where Boolean values are represented by a flow pressure in the channels.  In Sect.~\ref{liquidmarbles} The liquid is discretised in droplets and liquid marbles and computation is implemented via liquid marbles colliding with each other or actuating mechanical devices.  Oxidation wave fronts or diffusing fronts of precipitation solve logical and computational geometry problems in the reaction-diffusion computers in Sect.~\ref{rectiondiffusioncomputers}.   

The examples discussed demonstrate versatility of the liquid as a computing substrate. We hope the overview will encourage researchers and engineers to be more proactive in employing the liquid phase in their unconventional computing devices.

\section{Hydraulic algebraic machines}
\label{hydraulicmachines}

\begin{figure}[!tbp]
    \centering
    \subfigure[]{\includegraphics[width=0.2\textwidth]{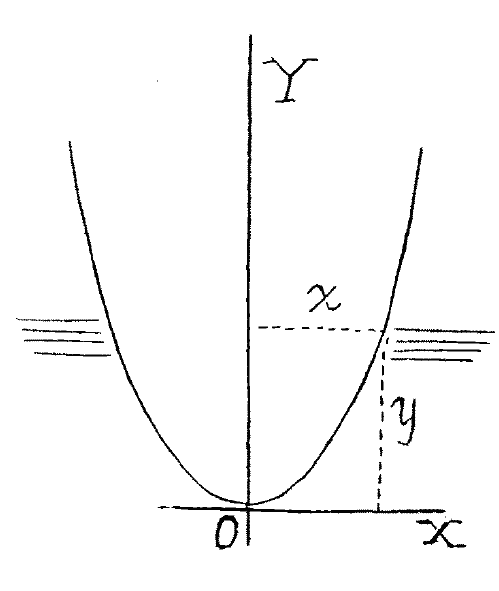}\label{emch1}}
    \subfigure[]{\includegraphics[width=0.2\textwidth]{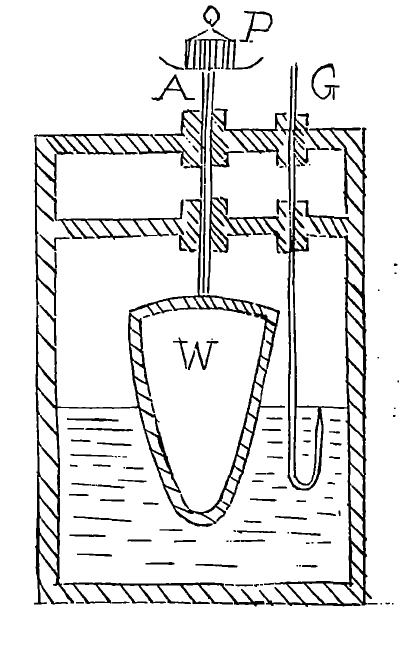}\label{emch2}}
    \subfigure[]{\includegraphics[width=0.24\textwidth]{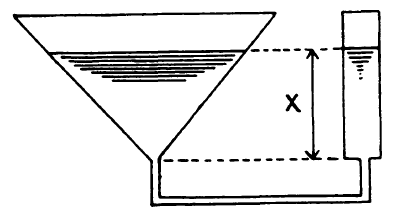}\label{dementwater1}}
    \subfigure[]{\includegraphics[width=0.3\textwidth]{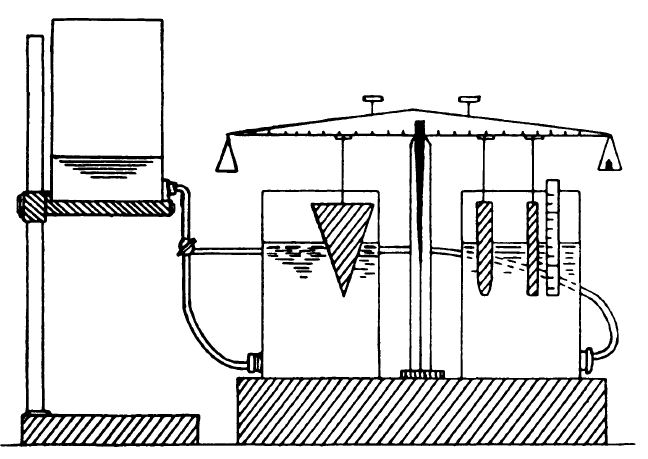}\label{meslinwater}}
    \caption{(ab)~Illustration of the hydraulic method for extracting $n$th root of any number proposed by Arnold Emch in 1901. From~\cite{emch1901two}.
    (cd)~Pictures from book 
    }
    \label{fig:rootextracrtor}
\end{figure}

In 1901 Arnold Emch published a paper in \emph{The American Mathematical Monthly} where he a proposed to calculate the $n$\textsuperscript{th} root of any number by immersing solid bodies in a liquid~\cite{emch1901two}. He assumed a paraboloid shape body is immersed in a liquid as shown in Fig.~\ref{emch1}. When this body is immersed a weight of the displaced water will be $W=\pi \cdot w \cdot \int_{0}^{y} x^2 dy=\pi \cdot w \cdot \int_{0}^{y} x^2 dy$, where $w$ is a weight of one cubic foot of water. The function $f(x)$ is selected such that the weight $W$ of the displaced water equals to $n$\textsuperscript{th} power of a given number $y$: $\pi \cdot w \cdot \int_{0}^{y} x^2 dy = y^n$. 

A device for the extraction of a square root of a given number $N$ is shown in Fig.~\ref{emch2}. $G$ is a hook-gauge showing a water level. The shape of the parabolic body immersed is defined by equation $y=x^2 \cdot \sqrt{\frac{\pi w}{2}}$. Let $W$ be a weight of the water displaced, $Q$ be a weight of the vessel, and $P$ be an added weight to make $Q \cdot P = N = y^2$. Thus, $y=\sqrt{N}$. The value $y$ is measured by gauge $G$ as a difference of the water levels  before and after the immersion. 

A method of solving trinomial equation $x^3+x=c$, where $c$ is a constant, was proposed by Demanet, cited by~\cite{gibb1914}. The device shown in Fig.~\ref{dementwater1} is an inverted cone joined with a cylinder, base is one square centimeter, with a tube. Height of the cone $H$ and the cone's radius $R$ are selected so that  $\frac{R}{H}=\frac{\sqrt{3}}{\sqrt{\pi}}$. A $c$~cm\textsuperscript{2} of water poured into the device will stay at the height $h$ in both vessels. Volume of the water in a cylinder will be $V=h^3$ (because $R=h\cdot \frac{\sqrt{3}}{\sqrt{\pi}}$). Volume of the water in the cylinder used in the device is $1 \cdot h$. Thus we have $h^3+h=c$. The height $h$ of the water is the solution of the equation. 

A hydrostatic balance, inspired by \cite{emch1901two}, was proposed by Meslin~\cite{gibb1914} to solve the equation $p\cdot x^m+ q\cdot x^n + \ldots =A$ (Fig.~\ref{meslinwater}~\cite{gibb1914}). Solid bodies are immersed in a liquid. Parameters of the bodies are selected so that when $x$ units of length are immersed in the liquid the volumes are proportional to $x^m, x^n, \ldots$. Coefficients $p$, $q$ etc. are represented by distances of the bodies' hanging points relative to the axis of rotation of the beam (left means negative and right positive). After bodies are immersed a weight $|A|$ is suspended at a unit distance from the axis of rotation. The system becomes temporarily disturbed but the equilibrium is restored by the water redistribured between the vessels. After that the moments of the bodies relative to the axis of rotation of the beam will be $p\cdot h^m + q\cdot h^n + \ldots$, and therefore $h$ is the solution found.\footnote{The hydraulic equation solvers are discussed in the context of electrical and mechanical solvers in \cite{frame1945machines}.}

\section{Hydraulic integrators}
\label{hydraulicintegrations}

\begin{figure}[!tbp]
    \centering
    %\subfigure[]{\includegraphics[width=0.3\textwidth]{EwbanksBookTItlePage}}
    \subfigure[]{\includegraphics[width=0.6\textwidth]{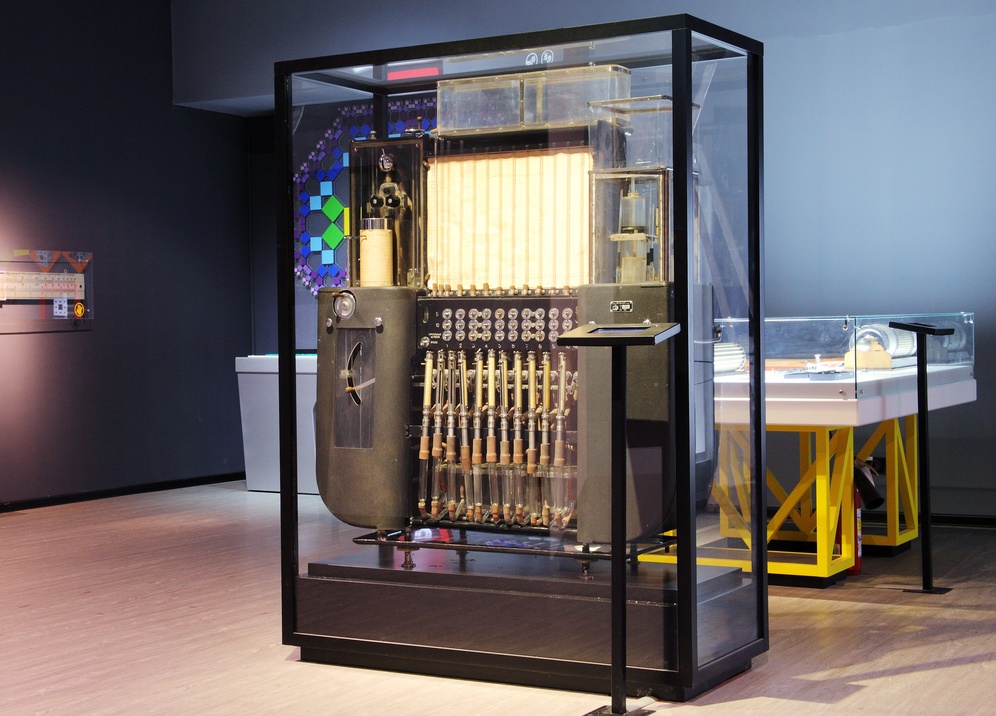}\label{LukyanovIntegrator}}
     \subfigure[]{\includegraphics[width=0.3\textwidth]{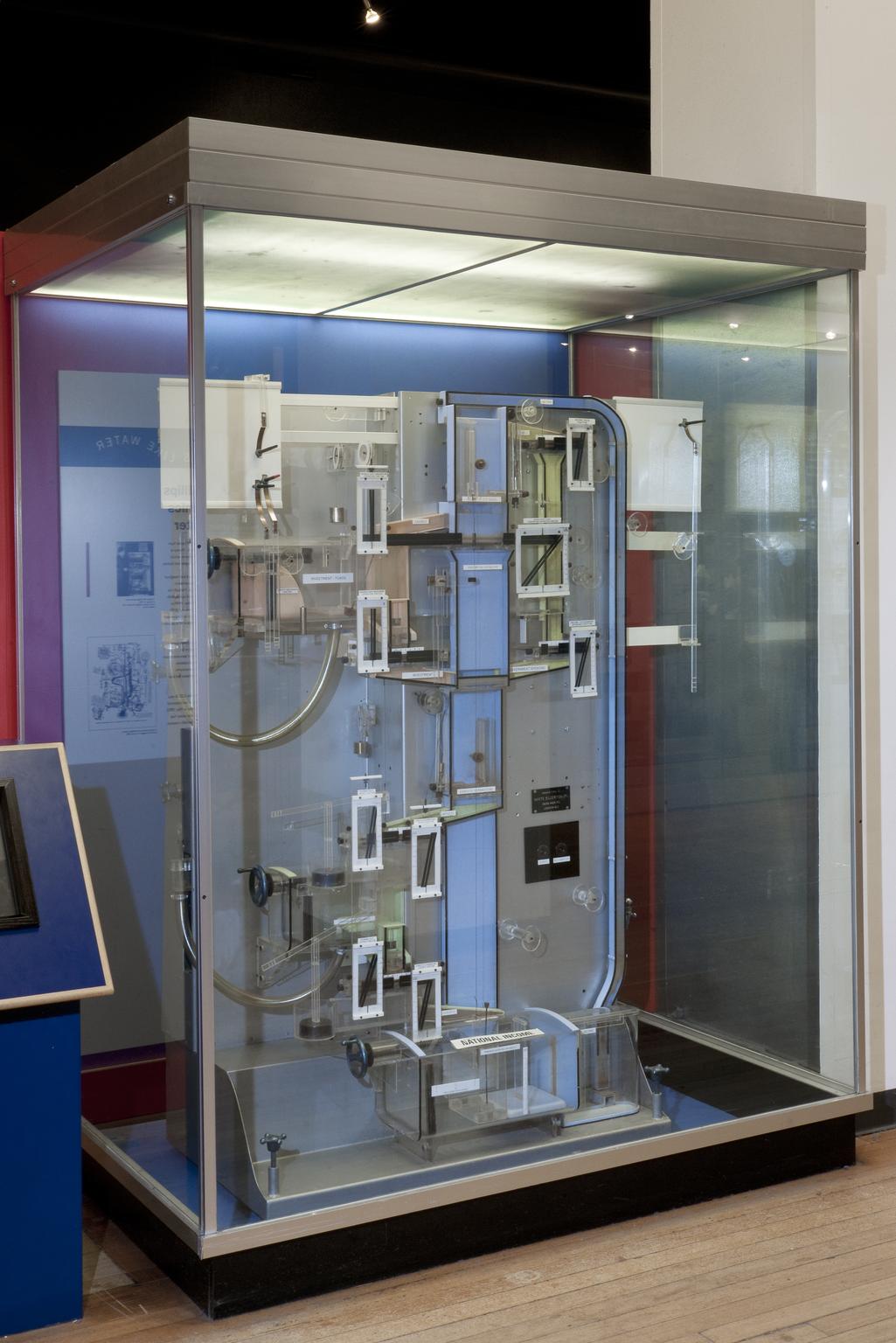}\label{PhillipsMachine}}
    \caption{
    %(a)~Title page of Ewbank's book on hydraulic machines (1842)~\cite{ewbank1842descriptive}.
   Hydraulic equation solvers.
    (a)~Lukyanov's Hydraulic Integrator (1936). This is a photo of Model No 3 Hydraulic integrator donated by Ryazan Factory of Analog Computers to Moscow Polytechnic Institute in 1956. Photo \copyright 2005-2018 Polytechnic Museum (Moscow, Russia).  
    (b)~Phillips Hydraulic Computer (1949). Science Museum Group Collection. \copyright The Board of Trustees of the Science Museum.
    }
    \label{fig:hydraulicintegrators}
\end{figure}

In early 1930s hydraulic computing devices have been invented simultaneously in USSR by Luk'yanov~\cite{luk1939hydraulic} and in USA by Moore~\cite{moore1936hydrocal}. Both were designed to imitate heath transfer not by solving differential equations by hand or existing calculators but by the analog modelling of the heat propagation with water (Fig.~\ref{LukyanovIntegrator}).  The devices relied on the following analogies between liquid and  thermal characteristics of thermoconductive building materials~\cite{moore1936hydrocal,luk1939hydraulic,arlazorov1952}. Levels of water in vessels represent difference of temperatures of the building materials and the air. Cut area of the vessels represents a thermal capacity of layers. Hydraulic resistance of  the tubes connecting the vessels is analogous to the thermal resistance of the simulate material layers.

As stated by Polytechnic Museum (Moscow, Russia)~\cite{moscowpolytech} there were c. 150 hydraulic integrators produced in USSR, some exported to Poland, Czech Republic and China. A portable version of Lukyanov integrator was manufactured for schools. Moore's hydrocal was deployed for studies on heat transfer in geological constructions.

Over a decade after the invention and relatively wide usage of the hydraulic integrators, in late 1940s, Phillips designed and prototyped his Monetary National Income Analogue Computer (MONIAC), also known as Phillips Hydraulic Computer (Fig.~\ref{PhillipsMachine}). There a flow of money was imitated by a redistribution of water between the containers~\cite{bissell2007historical}.

\section{Fluid mappers and maze solvers}
\label{fluidmappersmazesolvers}

In 1900 Hele-Shaw and Hay proposed an analogy between stream lines of a fluid flow in a thin layer and the lines of magnetic induction in a uniform magnetic field~\cite{hele1900lines}. They applied their ideas to solve  a ``problem of the magnetic flux distortion brought about by armature teeth''~\cite{hele1905hydrodynamical}. In 1940s Hele-Shaw and Hay analogies were advanced by Arthur Dearth Moore who developed fluid flow mapping devices~\cite{moore1949fields}.

\begin{figure}[!tbp]
\centering
\subfigure[]{\includegraphics[width=0.4\textwidth]{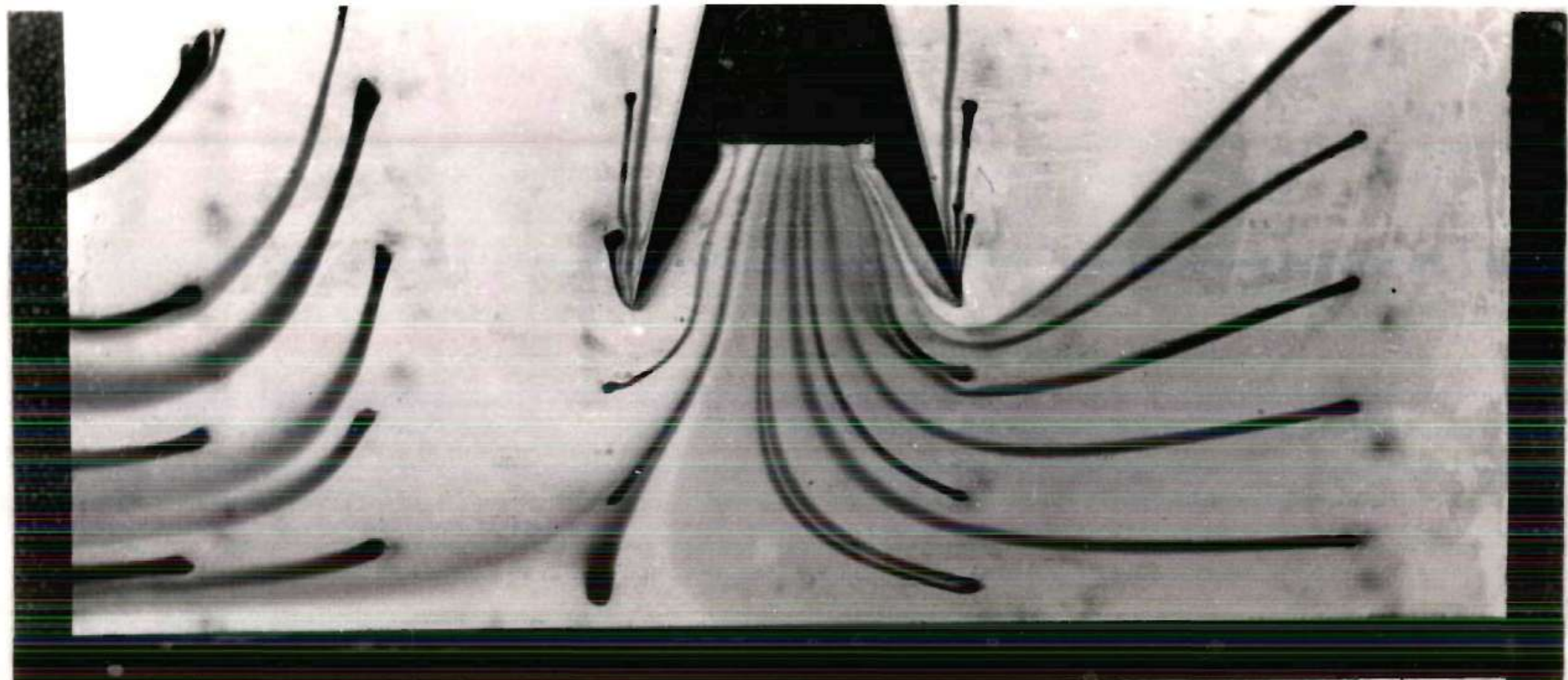}}
\subfigure[]{\includegraphics[width=0.25\textwidth]{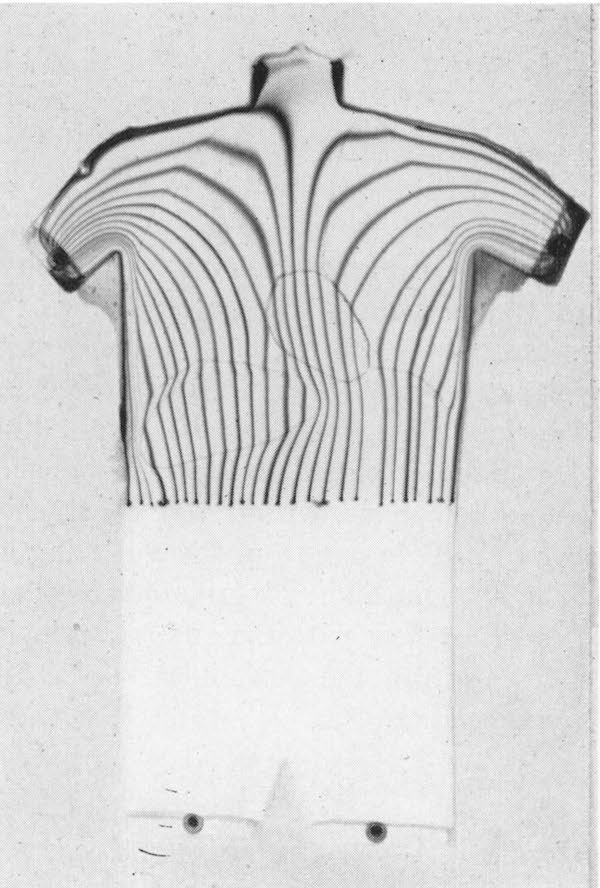}}
\subfigure[]{\includegraphics[width=0.3\textwidth]{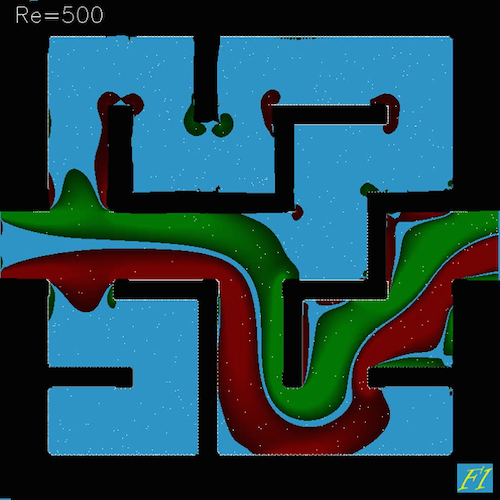}}
\subfigure[35 sec]{\includegraphics[width=0.35\textwidth]{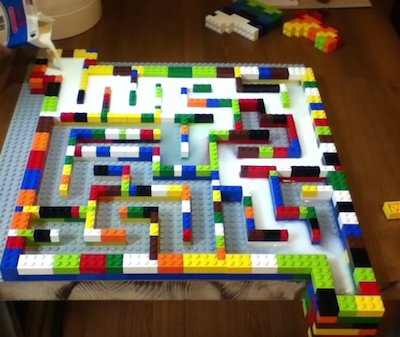}}
\subfigure[]{\includegraphics[width=0.3\textwidth]{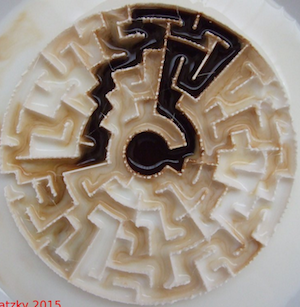}}
\caption{Fluid mappers and solvers.
(a)~ A fluid mapper used in optimisation of a canopy exhaust hood in 1954~\cite{clem1954use}.
(b)~An Imitation of the current flow in a human body with a thin-layer fluid flow, with domains of the low permeability corresponding to the lungs and liver, the fluid enters the model from the left leg and leaves the model through the arms. The experiments are conducted in 1952.
From~\cite{mcfee1952graphic}.
(c)~ The fluid flow through the maze. The entrance is on the left, the exit is on the right. The flow simulation is done in Flow Illustrator \protect\url{http://www.flowillustrator.com/} for visual flow control $dt=0.01$ and Reynolds number 500. Maze is black, red coloured areas are parts of fluid making clockwise rotation and green coloured areas --- conter-clockwise.
(d)~Maze solving with milk and water; snapshots from the video of experiments by Masakazu Matsumoto \protect\url{https://youtu.be/nDyGEq_ugGo} with kind permission from Masakazu Matsumoto.
(e)~Labyrinth solving with coffee and milk: the path is traced by the coffee~\cite{adamatzky2017physical}.
}
\label{fluid}
\end{figure}

The Moore's fluid mappers were made of a cast slab, covered by a glass plate, with input (source) and output (sink) ports. Crystals of  potassium permanganate  or methylene blue were evenly distributed at the bottom of the slab. Fluid flow lines were visualised by traces from the dissolving crystals. Moore shown that the fluid mappers can simulate electrostatic and magnetic fields, electric current, heat transfer and chemical diffusion~\cite{moore1949fields}. The fluid mappers became popular, for a decade, and have been used to solve engineering problems of underground gas recovery and canal 
seepage~\cite{MoorePages}, current flow modelling in human body~\cite{mcfee1952graphic} (Fig.~\ref{fluid}a) and design of fume exhaust hoods~\cite{clem1954use} (Fig.~\ref{fluid}b).

The Moore's fluid mappers could solve mazes, however Moore never reported this. A first published evidence of an experimental laboratory fluid maze solver is dated back to 2003. In the  fluidic maze solver developed in \cite{fuerstman2003solving} a maze is the network of micro-channels. The network is sealed. Only the source site (inlet) and the destination site (outlet) are open. The maze  is filled with a high-viscosity fluid. A low-viscosity coloured fluid  is pumped under pressure into the maze, via the inlet. Due to a pressure drop between the inlet and the outlet liquids start leaving the maze via the outlet. A velocity of the fluid in a channel is inversely proportional to the length of the channel. High-viscosity fluid in the channels leading to dead ends prevents the coloured low-viscosity fluid from entering the channels.  The shortest path --- least hydrodynamic resistance  path ---  from the inlet to the outlet is represented by the channels filled with coloured fluid (Fig.~\ref{fluid}c). Similar approach could be used to solve maze at a macro scale, e.g. with milk and water (Fig.~\ref{fluid}d) or milk and coffee (Fig.~\ref{fluid}e).

\section{Droplets tracing fluid mappers}
\label{mazesolvingdroplets}

A motion of a droplets of one liquid in another liquid is determined by thermal and chemical gradients, and directed by flows outside and inside the droplet~\cite{young1959motion,levich1962physicochemical,golovin1986chemothermocapillary,golovin1990drift,velarde1998drops,yoshinaga2014spontaneous}. If the gradients represent a solution of a computational problem then droplets travelling along the gradients might be seen as solving the problem. Such an experimental prototype of a droplet traversing pH gradient is presented in~\cite{lagzi2010maze}. There a polydimethylsiloxane maze is  is filled with a solution of potassium hydroxide. Surfactant is added to reduce the liquids surface pressure. An agarose block soaked in a hydrochloric acid  is placed at a destination site. A pH gradient establishes in the maze. Then a droplet of a mixture of mineral oil or dichloromethane with 2-hexyldecanoic acid is placed at the start site. The droplet is not mixed with solution filling the channels. The droplet travels along the steepest gradient of the potassium hydroxide. The steepest gradient is along the shortest path. Therefore the droplet travels from its start site to the destination site along the shortest path~\cite{lagzi2010maze}.  

Exact mechanics of the droplet motion is explained in~\cite{lagzi2010maze} as follows. Potassium hydroxide, which fills the maze, is a deprotonating agent. Molecules of the potassium hydroxide remove  protons from molecules of  2-hexyldecanoic acid diffusing from the droplet. A degree of the protonation is proportional to the concentration of hydrochloric acid diffusing from the destination site. The protonated 2-hexyldecanoic acid at the liquid surface determines the surface tension. The gradient of the protonated acid determines a gradient of the surface tension.  The surface tension decreases towards the destination site. A flow of liquid --- the Marangoni flow ---  is established from the site of the low surface tension to the site of the high surface tension. The droplet is moved by the flow~\cite{lagzi2010maze}. The flow of liquid between two sites of the maze have been also visualised with A dye powder, Phenol Rd, placed as the start site~\cite{lovass2015maze}. The Marangoni flow transports the dye form the start to the destination. The coloured channels represent a path connecting the start and the destination. This type of visualiation has been already in connection with Moore's fluid mappers in Sect.~\ref{fluidmappersmazesolvers}.

\begin{figure}[!tbp]
    \centering
    \subfigure[0~s]{\includegraphics[scale=0.4]{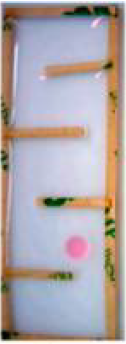}}
     \subfigure[20~s]{\includegraphics[scale=0.4]{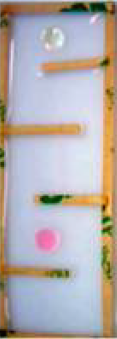}}
      \subfigure[60~s]{\includegraphics[scale=0.4]{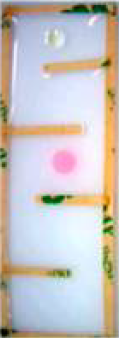}}
       \subfigure[140~s]{\includegraphics[scale=0.4]{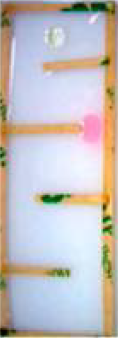}}
        \subfigure[180~s]{\includegraphics[scale=0.4]{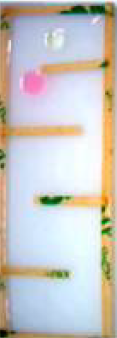}}
        \subfigure[200~s]{\includegraphics[scale=0.4]{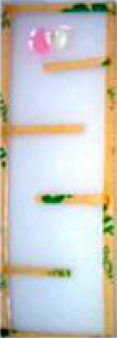}}
    \caption{A decanol droplet solves maze. From~\cite{cejkova2014dynamics} with kind permission of Jitka \v{C}ejkova.}
    \label{fig:JitkaDroplet}
\end{figure}

Another prototype of chemotactic droplet maze solver is demonstrated in \cite{cejkova2014dynamics}. The maze is filled with a water solution of a sodium decanoate. A nitrobenzene droplet loaded with sodium chloride is placed in the destination site. A  decanol droplet is placed at the start site. The sodium chloride diffuses from the its host nitrobenzene droplet in the destination site. A gradient of the saline concentration is established. The gradient is steepest along a shortest path leading from any site of the maze to the destination. The decanol droplets move along the steepest gradient till it reaches the droplet at the destination site (Fig.~\ref{fig:JitkaDroplet}). 

The experimental prototypes of travelling droplets laid the foundation of an emergent field of liquid robots~\cite{vcejkova2017droplets,chiolerio2017smart}. There droplets are studied as adaptive actuating and/or propulsive devices capable for navigation in a dynamically changing environments. 

While talking about mobile droplets we must mention that liquid droplets of Belousov-Zhabotinsky reaction (Sect.~\ref{rectiondiffusioncomputers}) are known to exhibit spontaneous motion~\cite{kitahata2005chemo,kitahata2011spontaneous,miyazaki2015coupling,suematsu2016oscillation} due to convention processes coupled with oxidation waves. However, we are not aware of any programmable information processing tasks being implemented by the mobile BZ droplets (a computation with stationary BZ droplets is discussed in Sect.~\ref{rectiondiffusioncomputers}).

\section{Fluidic logic}
\label{fluidlogic}

Fluids have been used for centuries to transmit force and energy in mechanical systems. 
First works on using fluids for computation are traced back to late 1950s early 1960s~\cite{kirshner2012design, moylan1968fluid, conway1971guide}. The basic principles of the 
fluidic devices are the laminar flow of a fluid, a jet interaction, a wall attachment and a vortex effect. The jet interaction is the phenomenon where fluid flows are arranged so that small opposing 
jets experience changes of direction which can be used as output signals. The wall attachment phenomenon is that the fluid attaches to a surface within a device and continues to flow over the surface until disturbed.  First devices designed and fabricated in 1960s included the beam deflection, turbulence, vortex and wall attachment amplifiers,  the {\sc and}, {\sc not}, {\sc or} and {\sc xor} logical elements, counters and shift registers. The fluidic devices have been used 
in jet sensing, programmable sequence control, flameproof equipment,  machine tools control,  
systems operating nuclear reactor coolant, servo-control in marine applications, missile and aircraft control, 
artificial heart-pump, lung ventilator~\cite{kirshner2012design, moylan1968fluid, conway1971guide}.

\begin{figure}[!tbp]
\centering
\subfigure[]{\includegraphics[width=0.49\textwidth]{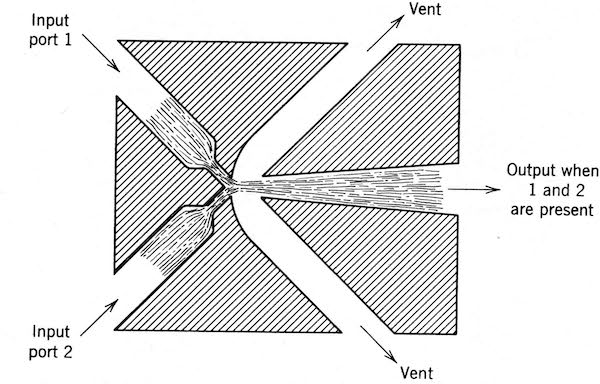}\label{AND412}}
\subfigure[]{\includegraphics[width=0.49\textwidth]{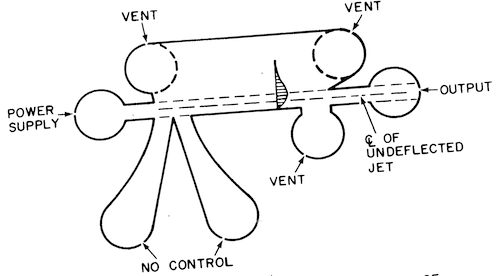}} 
\subfigure[]{\includegraphics[width=0.49\textwidth]{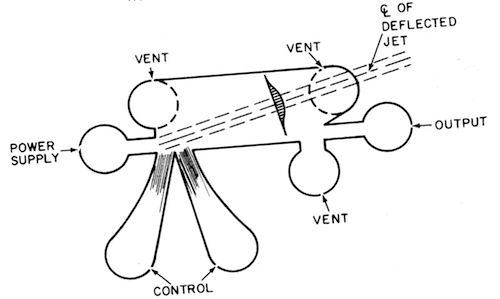}} 
\subfigure[]{\includegraphics[width=0.2\textwidth]{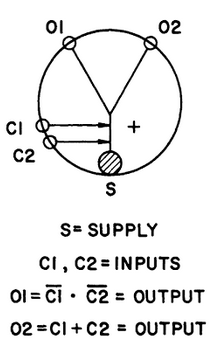}}
\subfigure[]{\includegraphics[angle=-90, scale=0.8]{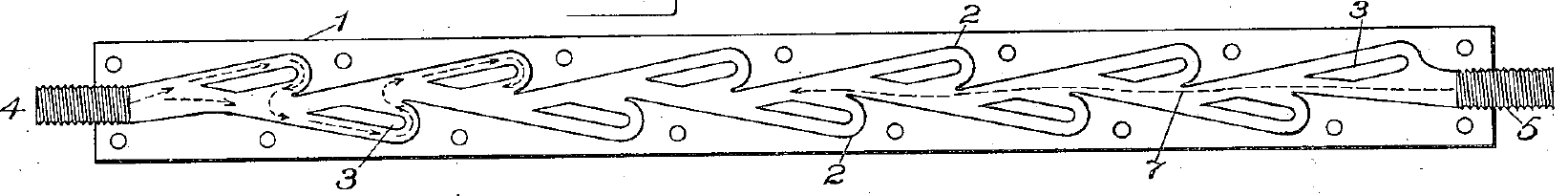}}
\subfigure[]{\includegraphics[scale=0.7]{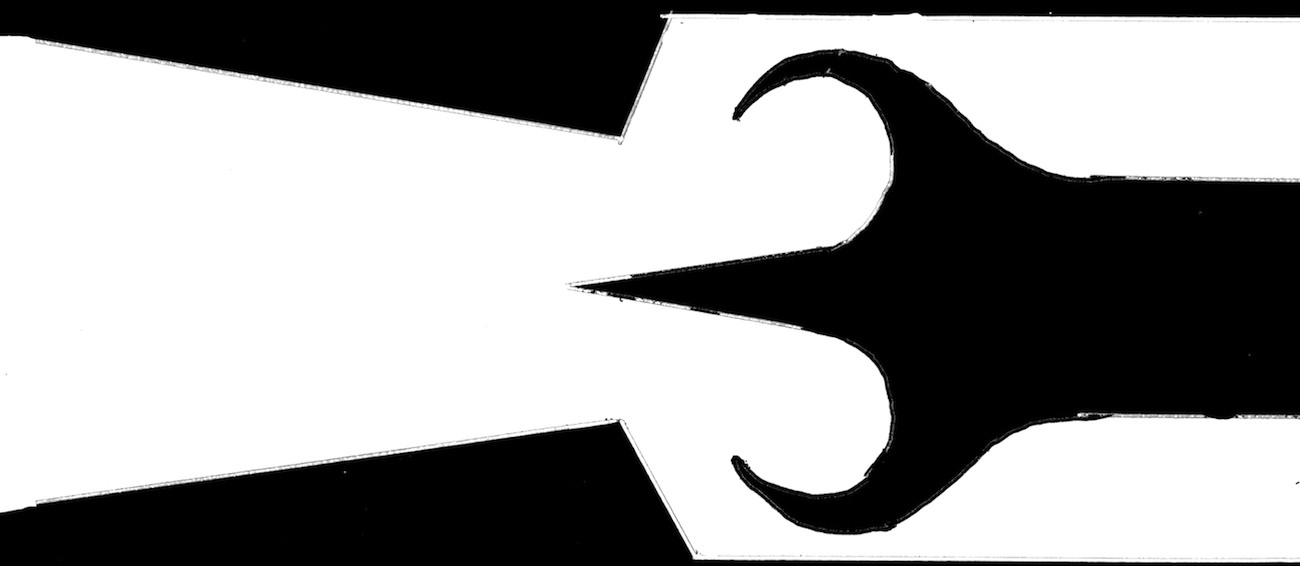}}
\subfigure[]{\includegraphics[scale=0.25]{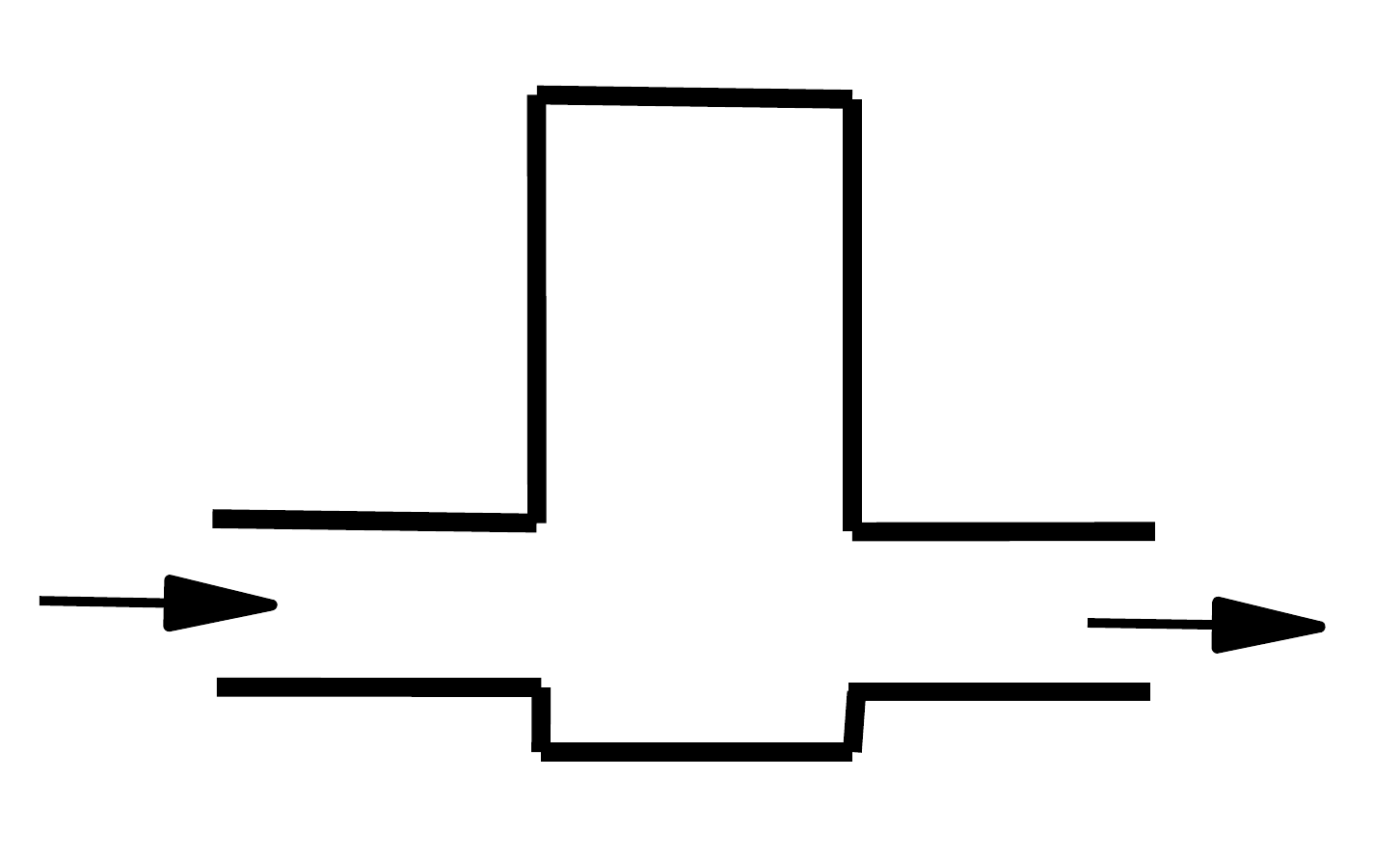}}
\subfigure[]{\includegraphics[scale=0.15]{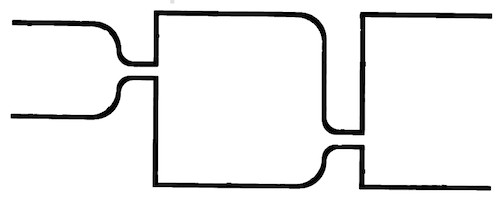}}
\subfigure[]{\includegraphics[scale=0.28]{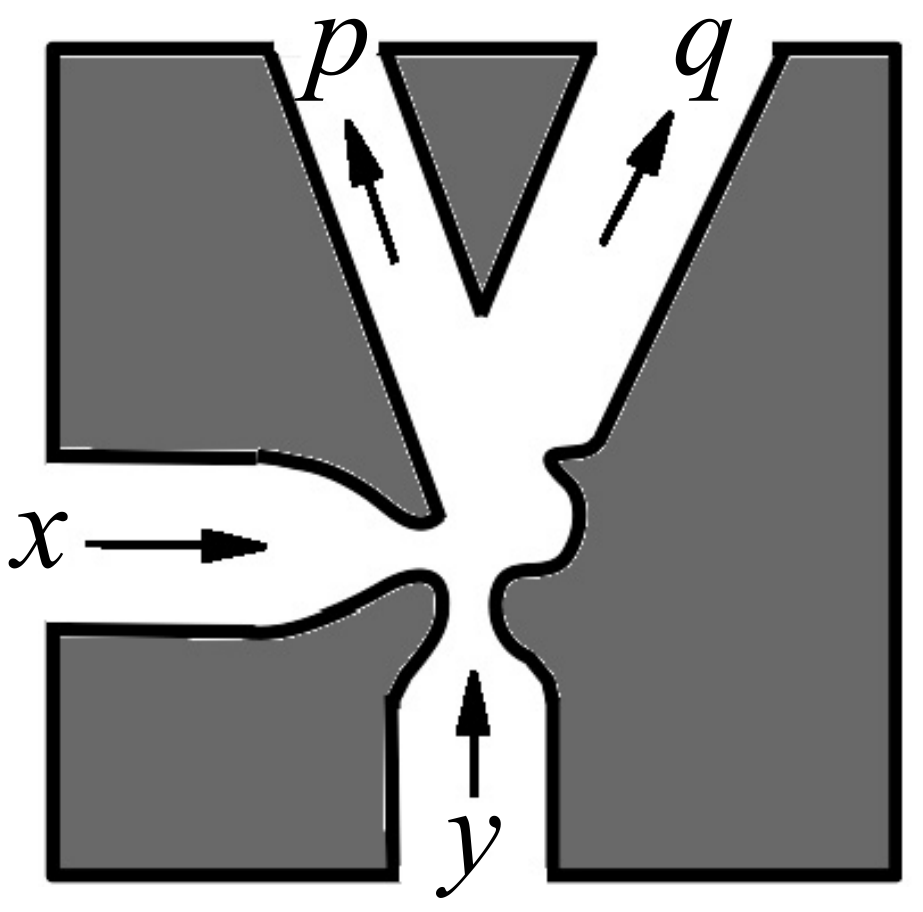}\label{HobbsGateScheme}}
\subfigure[]{\includegraphics[scale=0.18]{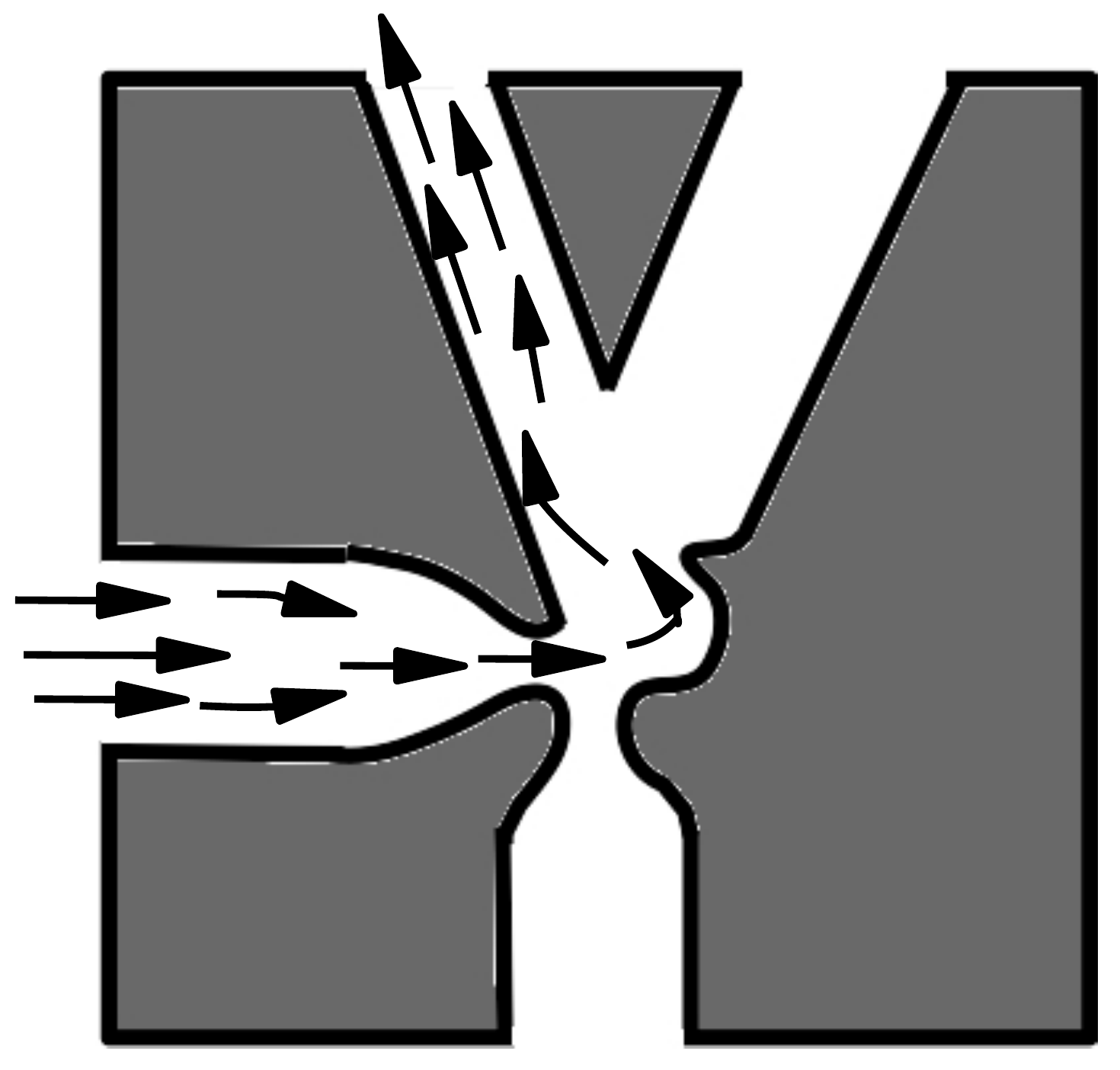}\label{Hobbs_X}}
\subfigure[]{\includegraphics[scale=0.18]{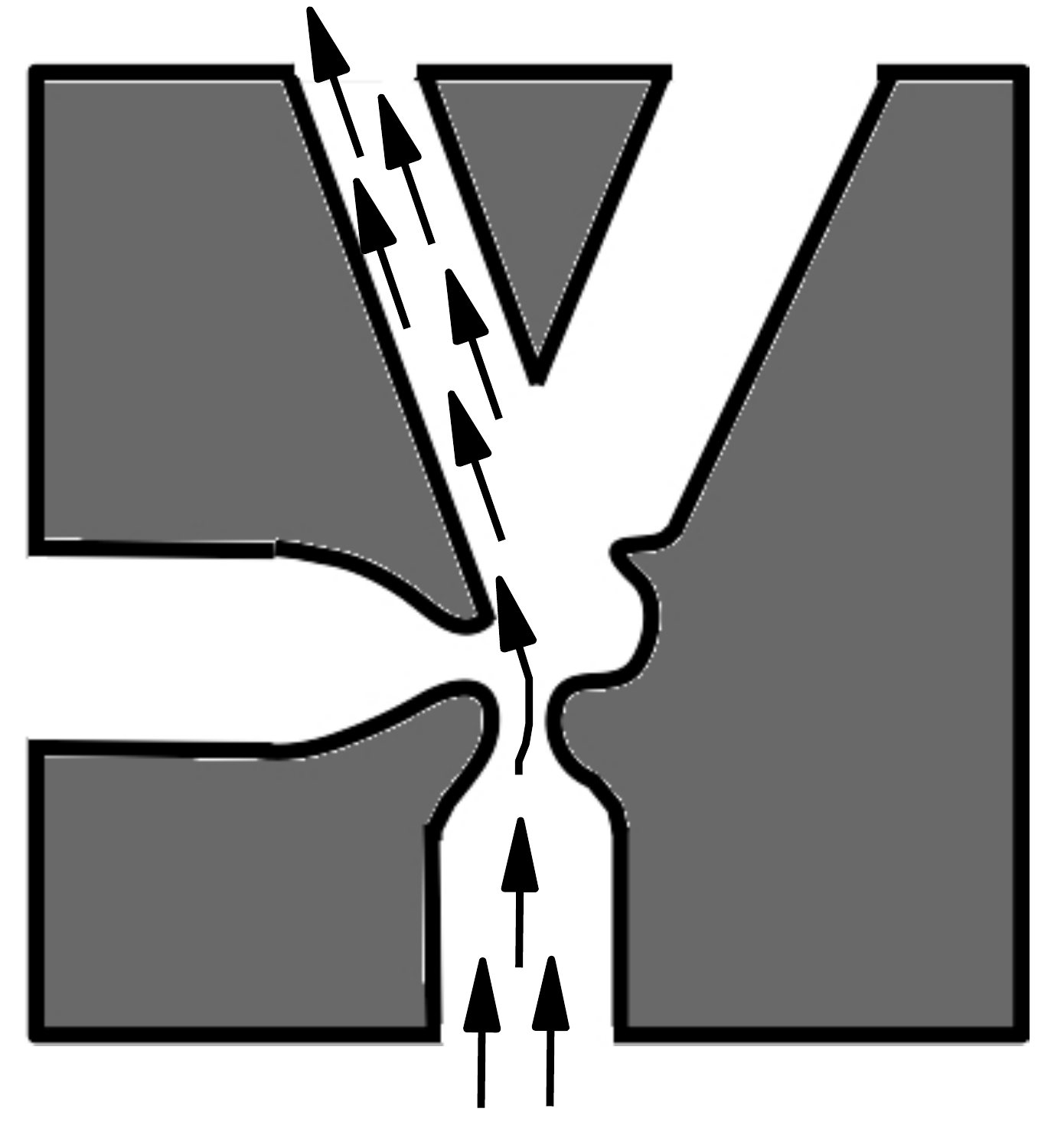}\label{Hobbs_Y}}
\subfigure[]{\includegraphics[scale=0.2=18]{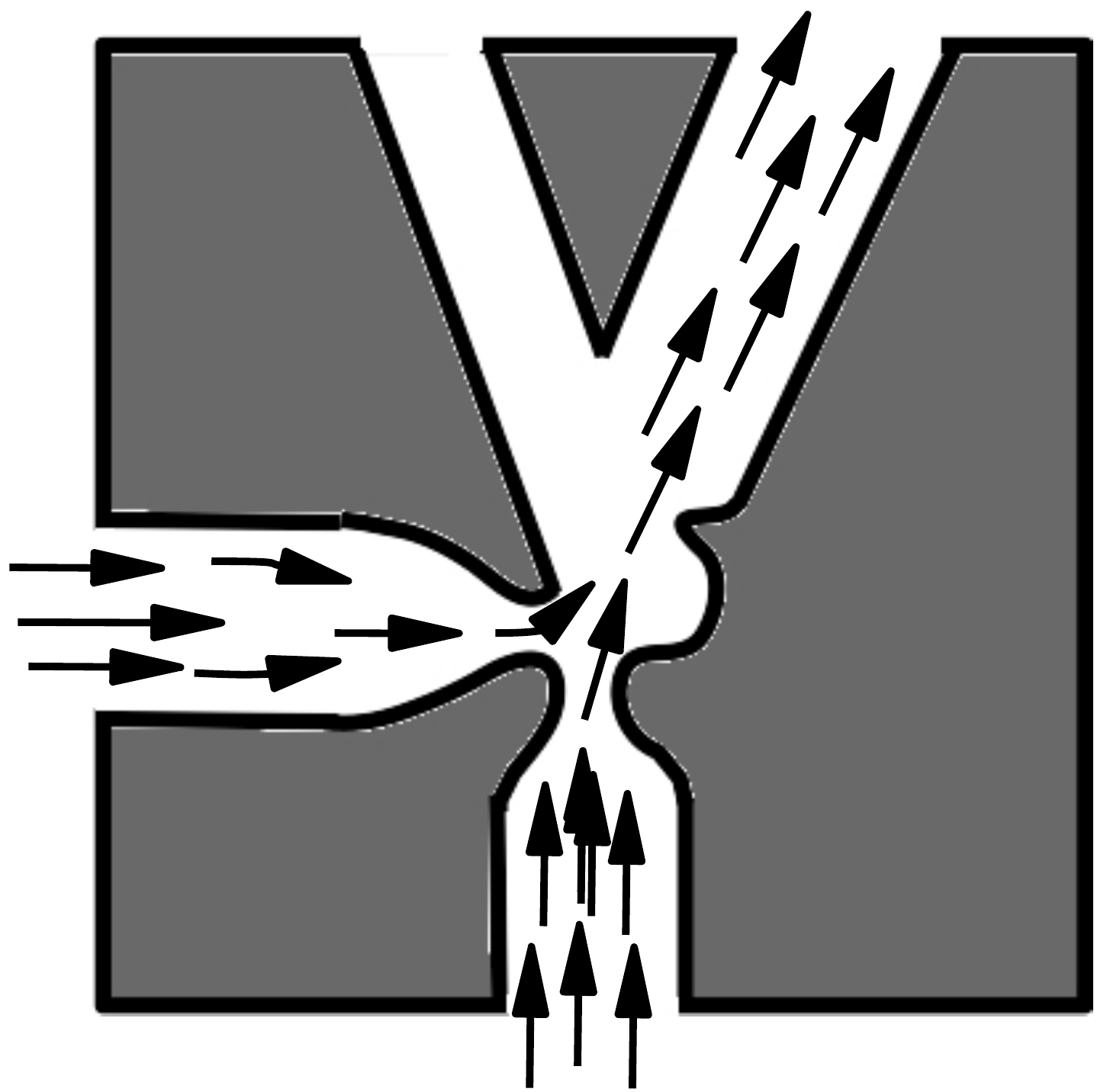}\label{Hobbs_XY}}
\caption{ 
(a)~Fluidic {\sc and-not} gate~\cite{peter1965and, belsterling1971fluidic}.
(bc)~Fluidic deflection type {\sc nor} element. From \cite{kirshner2012design}: (b)~non-deflected jet, (c)~jet deflected by control stream.
(d)~A diagram of the monostable fluid {\sc nor-or} amplifier~\cite{dummer2013fluidic}.
(e)~Tesla diode~\cite{tesla1920valvular}.
(f)~Scroll diode~\cite{paul1969fluid, kirshner2012design}.
(g)~Basic delay.  From~\cite{drake1965pure}.
(h)~Delay and diode. From~\cite{bowles1965passive}.
(i--l)~Hobbs gate:  a hook type fluidic half-adder
(i)~Structure of the gate 
(jkl)~Dynamics of the fluid streams for inputs (j)~$x=1$ and  $y=0$, (k)~$x=0$ and $y=1$, (l)~$x=1$ and $y=1$. 
Modified from \cite{hobbs1963fluid}.
}
\label{gates}
\end{figure}

The {\sc and-not} gate is the most known, a par with a bistable amplifier, devices in the fluidics  (Fig.~\ref{gates}a).  
Two nozzles are placed at right angles to each other.  
When there are jet flows in both nozzle they collide and merge  into a single jet entering the central outlet.  If the jet flow is present only in one of the input nozzles it goes into the vent. Horizontal output channel implements $x$ {\sc and} $y$ and later channels $x$ {\sc and not} $y$ and {\sc not} $x$ {\sc and} $y$.

A monostable beam deflection device is comprised of a power supply,  controls/inputs and vents (Fig.~\ref{gates}bc).
When no inputs are present the fluid jet from the power source exits through the output (Fig.~\ref{gates}b). 
When one or both input jets are present, the  jet from the power source is deflected into the vent and discharged 
(Fig.~\ref{gates}c)~\cite{kirshner2012design}. The fluid jet exits the output only if none of the input jets are present. 
This is {\sc nor} operation.

The monostable beam deflection device (Fig.~\ref{gates}bc) can be transformed into {\sc nor-or} gate (Fig.~\ref{gates}d) by adding an output outlet instead of the vent~\cite{dummer2013fluidic}.  When no control jets are present the jet from the power source exits via the outlet $O1$. If one or both signal jets are present, the jet from the power source is deflected in the outlet $O2$.

A fluidic diode is a two-terminal device which restricts, or even cancels, flow in one direction (backward direction). 
Tesla diode~\cite{tesla1920valvular} (Fig.~\ref{gates}e) and scroll diode~\cite{paul1969fluid, kirshner2012design} (Fig.~\ref{gates}f)  are most known fluidic diodes (as well as vortex diode which is not discussed here). 

The Tesla diode (Fig.~\ref{gates}e), called the `valvular conduit' by its inventor~\cite{tesla1920valvular}, is composed of 
buckets and partitions arranged in such a manner that the forward flow propagates mainly along axis (4 to 5 in Fig.~\ref{gates}e). In the backward direction (5 to 4 in Fig.~\ref{gates}e) fluid enters the branches and loops around to oppose the main flow.

In the fluidic scroll diode (Fig.~\ref{gates}f) the channel, or nozzle, is converging in the backward direction and enters an annular cap. In the forward direction the fluid flows through the throat and into a diffuser section. In the backward direction the fluid enters the cap and is directed back towards an incoming flow, causing a turbulence.

A delay in the fluidic systems is implemented as volumetric tank (Fig.~\ref{gates}g) with input and output pipes. 
A step change in the input pressure on the input appears as a similar change in the output pressure on the output after a delay.  The delay is caused by the turbulence. The amount of the delay is determined by the volume of the tank~\cite{drake1965pure}. Another version of a delay element  (Fig.~\ref{gates}h) combines orifices and volumes to have a low impedance in one direction of the flow 
(from the left to the right in  Fig.~\ref{gates}h) and a high impedance in the opposite direction of the flow (from the right to the left)~\cite{bowles1965passive}.
The impedance provides a phase shift during transient flow. The phase shift contributes to the retarding of the fluid flow, thus introducing a time delay of the flow.
The device can also act as a diode and a pressure divider.

A fluidic one-bit half-adder can be implemented on the basis of the gate proposed by Hobbs in 1963~\cite{hobbs1963fluid} (Fig.~\ref{HobbsGateScheme}). 
Logical values are encoded into presence of streams at the specified channels. When only input is {\sc True}, $x=1$ a power jet stream enters the gate via channel $x$. 
The stream  is turned by the hook and follows the channel $p$ (Fig.~\ref{Hobbs_X}). 
When only input $y=1$ the  power jet stream entering the gate via channel $y$ locks on (gets attached to) the left boundary wall
of its channel, and propagates along the channel $p$ (Fig.~\ref{Hobbs_Y}). 
If both inputs are {\sc True} streams entering $x$ and $y$ merge and follow the channel $q$ (Fig.~\ref{Hobbs_XY}). 
Thus the stream exiting the channel $p$ represents  $p = x \oplus y$ and the stream exiting the channel $q$ represents $q = xy$. 

\section{Digital microfluidic logic}

\begin{figure}[!tbp]
    \centering
    \subfigure[]{\includegraphics[scale=0.15]{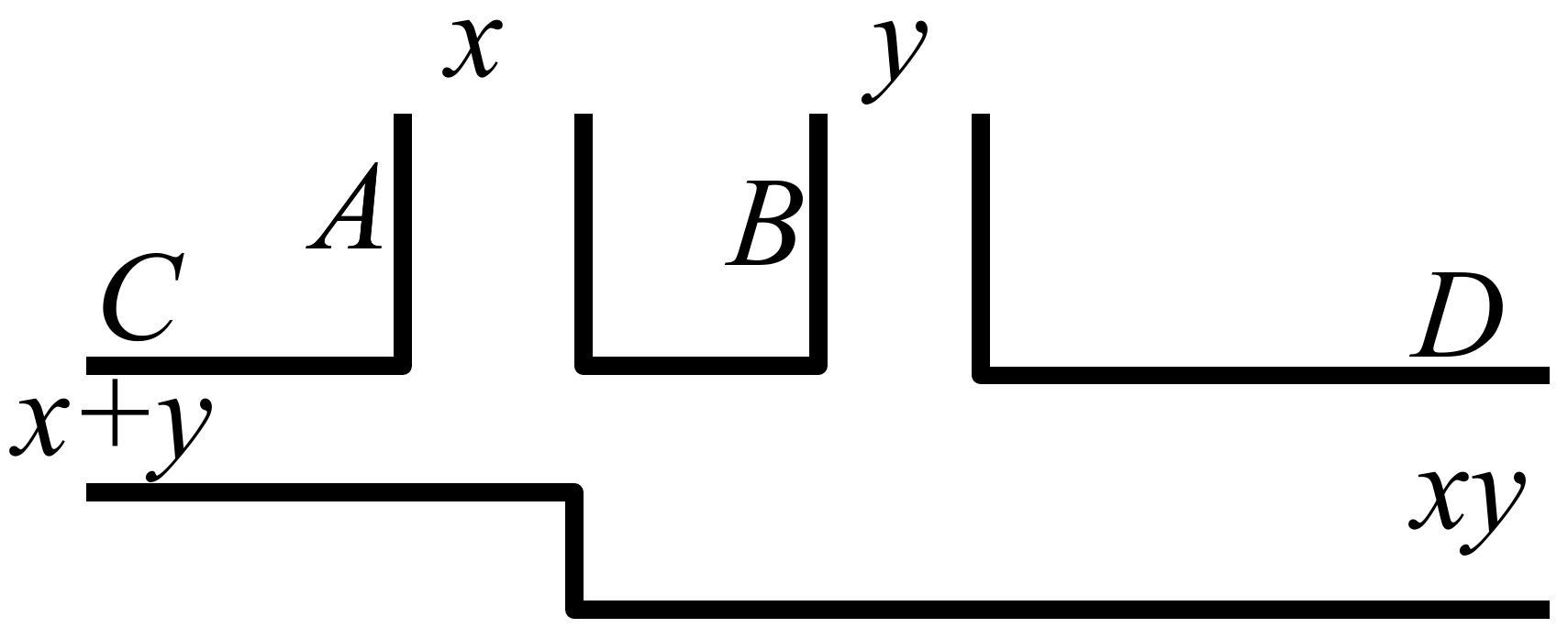}\label{fluidicgate_scheme}}
    \subfigure[]{\includegraphics[scale=0.15]{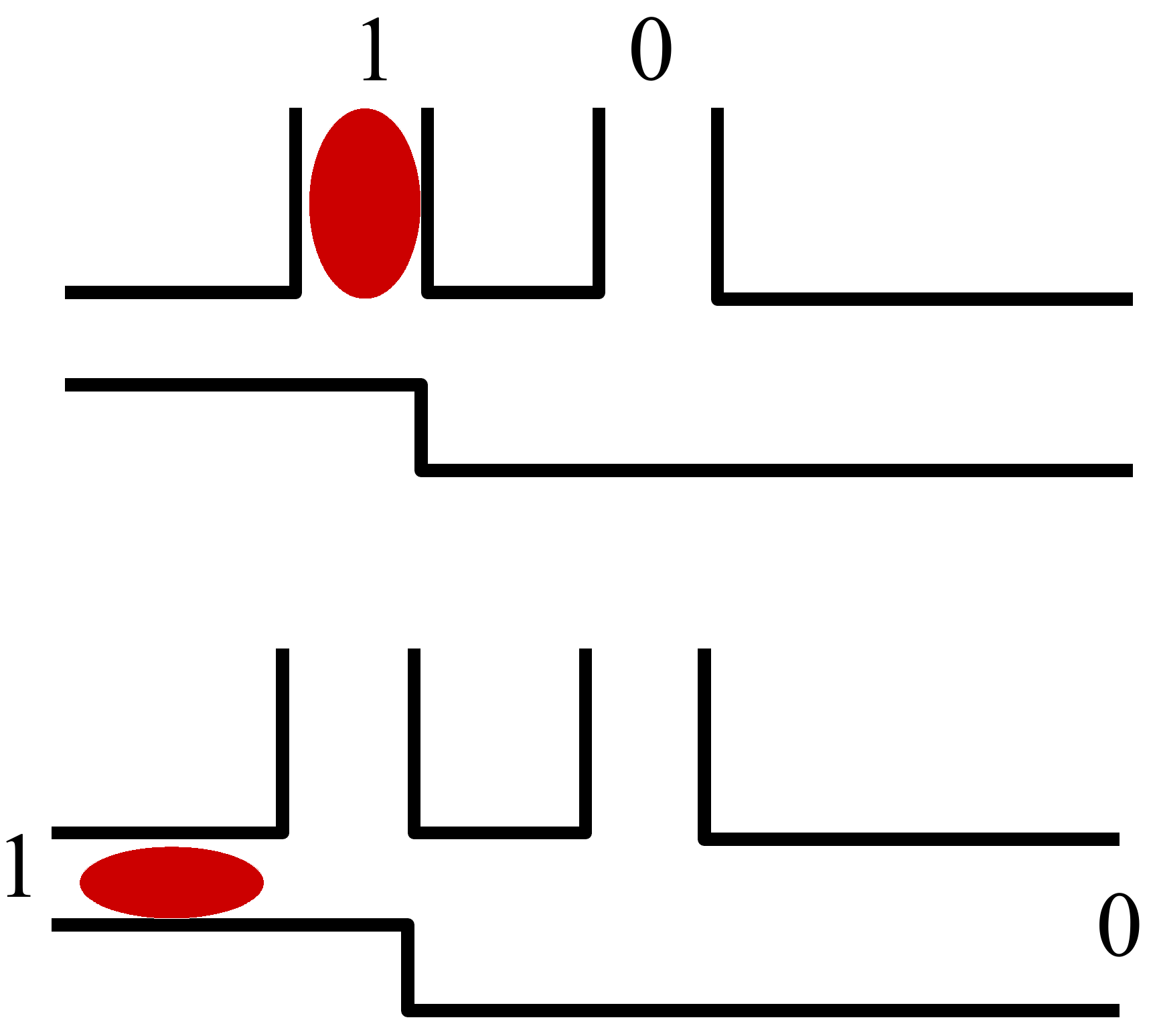}\label{fluidicgate_10}}
    \subfigure[]{\includegraphics[scale=0.15]{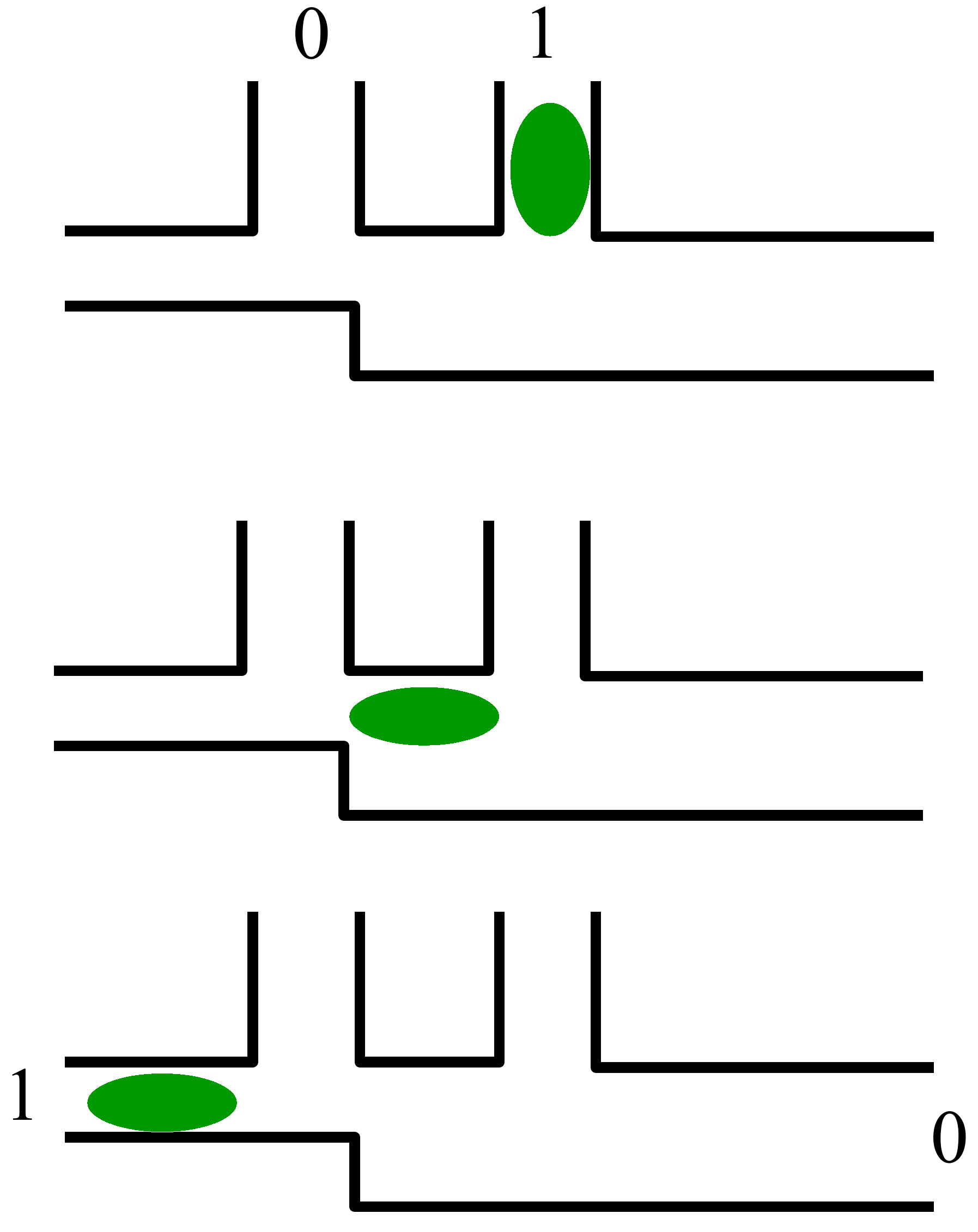}\label{fluidicgate_01}}
    \subfigure[]{\includegraphics[scale=0.15]{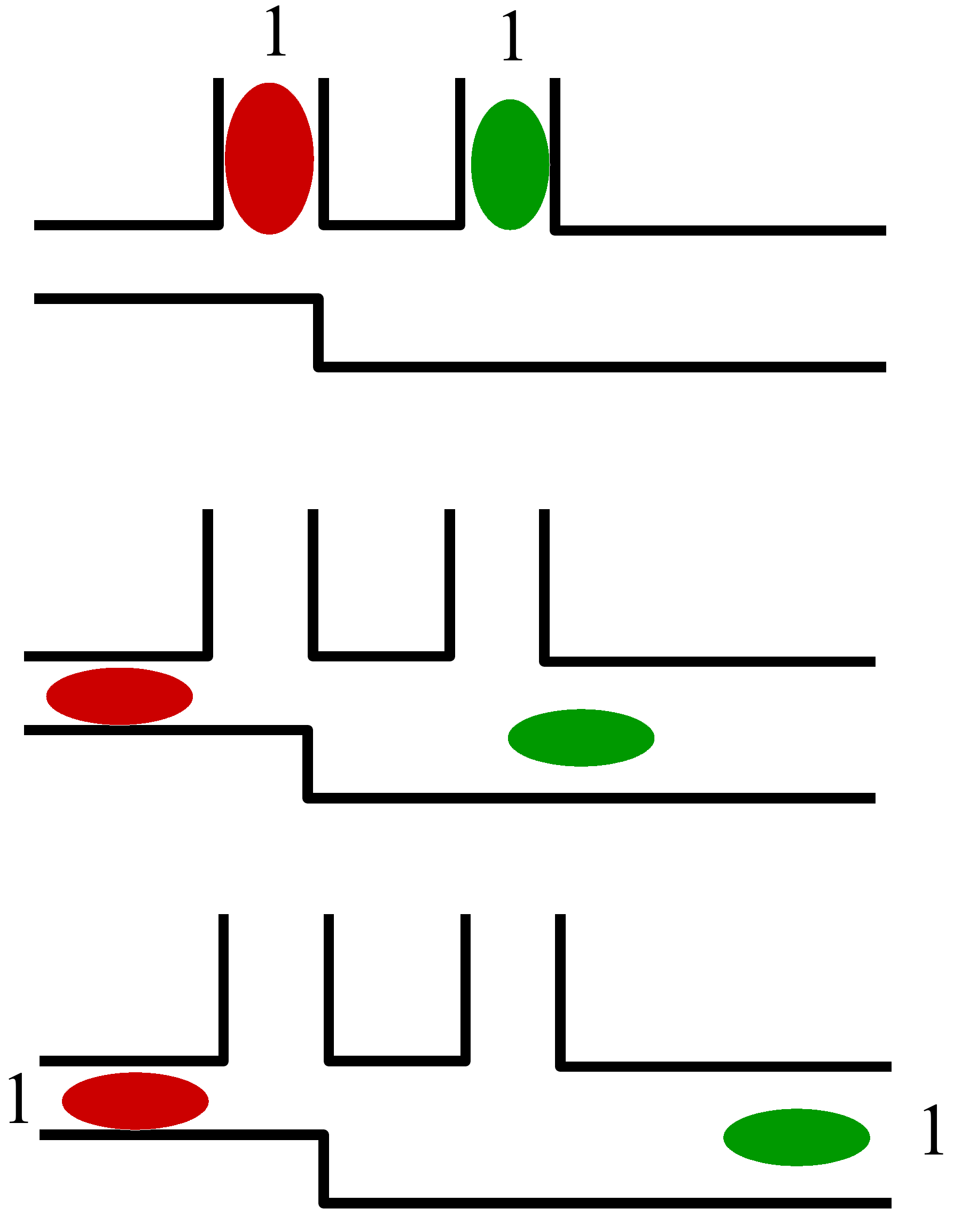}\label{fluidicgate_11}}
       \subfigure[]{\includegraphics[scale=0.35]{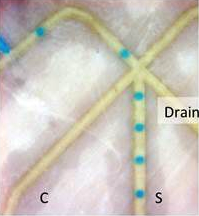}\label{Morgan_C}}
    \subfigure[]{\includegraphics[scale=0.35]{Morgan_AA}\label{Morgan_A}}
       \subfigure[]{\includegraphics[scale=0.35]{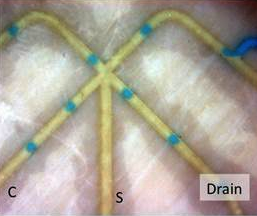}\label{Morgan_B}}
    \subfigure[]{\includegraphics[scale=0.35]{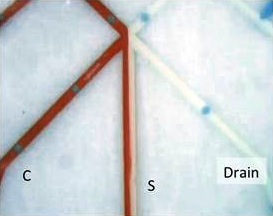}\label{Morgan_D}}
    \caption{Microfluidic gate {\sc or}-{\sc and}, redrawn from~\cite{cheow2007digital}; oil droplets are coloured red and green for distinction only.
    (e--d) Snapshots of experimental implementation of microfluidic half-adder  implemented by Morgan et al.~\cite{morgan2016simple}; (e) $x=1$, $y=0$, (f) $x=0$, $y=1$,  (g) $x=1$, $y=1$, (h) oil entering channel $x$ is coloured to demonstrate a deflection of flow. From ~\cite{morgan2016simple}.}
    \label{fig:fluidicdropletgate}
\end{figure}

In fluidic logic devices fluid jets are programmed by geometry of channels and perform computation by interacting with each. Boolean values are represented by pressure of the jets in the output channel. In digital microfluidcs signals are represented by droplets or bubbles travelling in the channel~\cite{fair2007digital,toepke2007microfluidic,mertaniemi2012rebounding,cheow2007digital}. The droplets/bubbles also control pressure in the channels thus affecting trajectories of other droplets/bubbles. Example a fluidic gate, designed in~\cite{cheow2007digital}, with two inputs and two outputs is shown in Fig.~\ref{fluidicgate_scheme}-\ref{fluidicgate_11}. 
An oil droplet is in aqueous phase. When a droplet enters a channel with the largest flow rate, a pressure drop increases across the channel containing two-phase emulsions, particularly at a low capillary
number.  

Cheow et al.~\cite{cheow2007digital} designed a ratio of tubes and flows (Fig.~\ref{fluidicgate_scheme}) such that the following phenomena take place. A flow  through $A$ to $C$ exceeds flow through $A$, the regime is laminar therefore all flow from $x$ goes via $AC$. An oil droplet  entering $A$ travels into $C$ (Fig.~\ref{fluidicgate_10}). A flow through the bridge between $A$ and $B$ exceeds half of the flow through $B$. Thus, an oil droplet entering $B$ also travels into $A$ 
(Fig.~\ref{fluidicgate_01}). When a droplet from $A$ enters channels $C$ a hydrodynamic resistance of $A$ increases. The flow via bridge connecting $A$ and $B$ becomes less than half of the flow through $B$. The droplet entering $B$ travels into channel $D$ (Fig.~\ref{fluidicgate_11}). 

Using similar principles, Morgan et al.~\cite{morgan2016simple} implemented one-bit half-adder. When only one of the input channels represent `1' the droplets from this channel goes into channel $S$ (Fig.~\ref{Morgan_C} and Fig.~\ref{Morgan_A}). When both inputs are `1' the droplets travel into lateral channels (Fig.~\ref{Morgan_B}). 

Other variants of fludic gates are realised in \cite{toepke2007microfluidic} using surface tension-based passive pumping and fluidic resistance, and in~\cite{prakash2007microfluidic}.

\section{Billiard ball computing with droplets and marbles}
\label{liquidmarbles}

Most logical gates implemented in fluidic devices (Sect.~\ref{fluidlogic}) and Belousov-Zhabotinsky chemical medium (Sect.~\ref{rectiondiffusioncomputers}) utilise the phenomenon of merging colliding jet streams (Fig.~\ref{AND412}) and excitation wave-fragments (Fig.~\ref{CollisionGate}). Thus, input jet streams or wave-fragments $x$ and $y$ propagating, beyond the collision site, along their original trajectories represent functions $x\overline{y}$ and $\overline{x}y$. A stream or a wave-fragment propagating along new trajectory represents the function $xy$.
Discrete soft and liquid bodies do not always merge when collide but often reflect. Thus they can implement collision-based gates.

A  collision-based computation, emerged from Fredkin-Toffoli conservative logic~\cite{fredkin2002conservative}, employs mobile compact finite patterns, which implement computation while interacting with each~\cite{adamatzky2002collision}. Information values (e.g. truth values of logical variables) are given by either absence or presence of the localisations or other parameters of the localisations. These localisations travel in space and perform computation when they collide with each other.  Thus the localisations undergo transformations, they change velocities, form bound state and annihilate or fuse when they interact with other localisations. Information values of localisations are transformed as a result of collision and thus a computation is implemented. 

\begin{figure}[!tbp]
\centering
\subfigure[]{\includegraphics[scale=0.27]{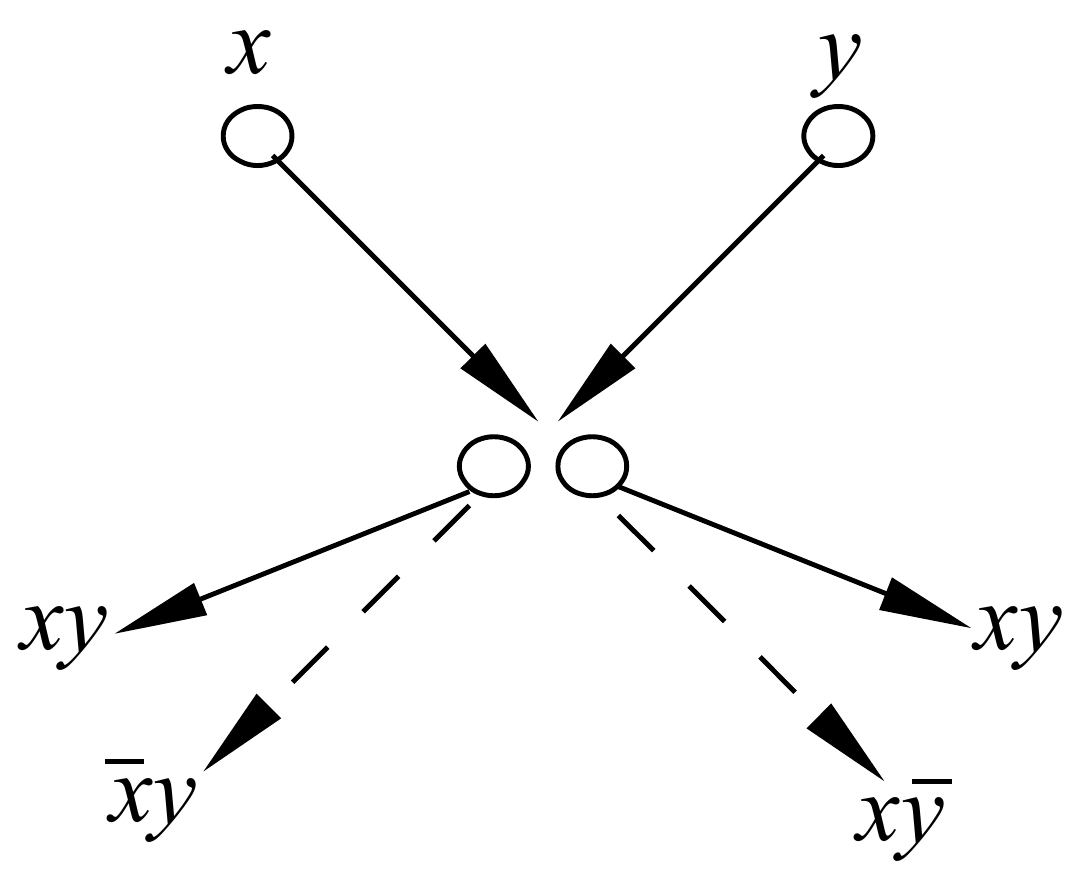}\label{HardBallsGate}}
\subfigure[]{\includegraphics[scale=0.27]{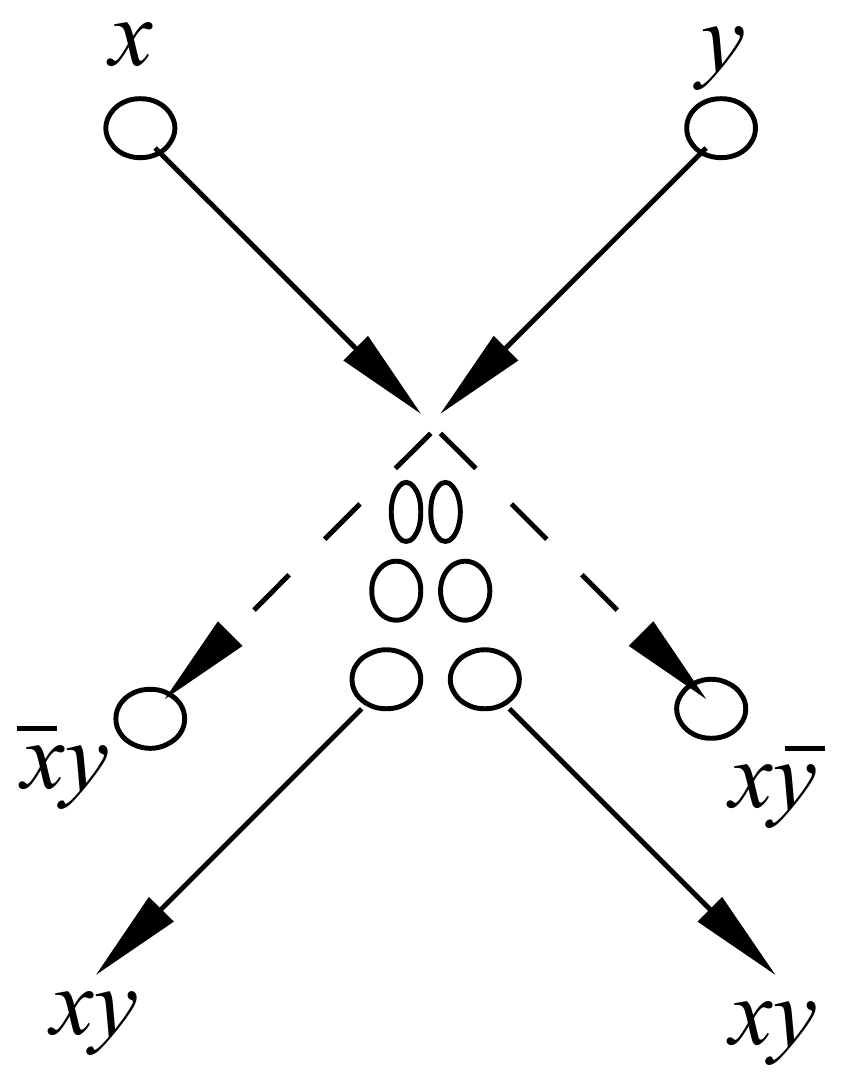}\label{SoftBallsGate}}
\subfigure[]{\includegraphics[scale=0.27]{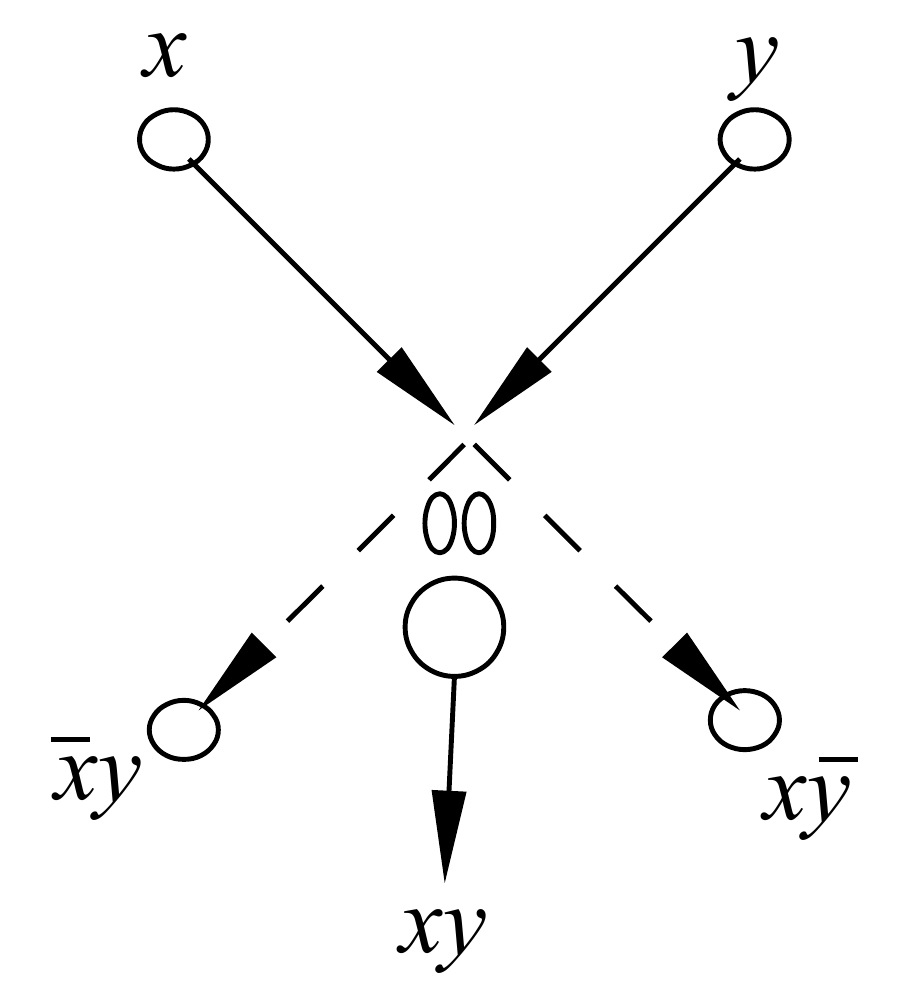}\label{MergingDropletsGate}}
\subfigure[]{\includegraphics[scale=0.22]{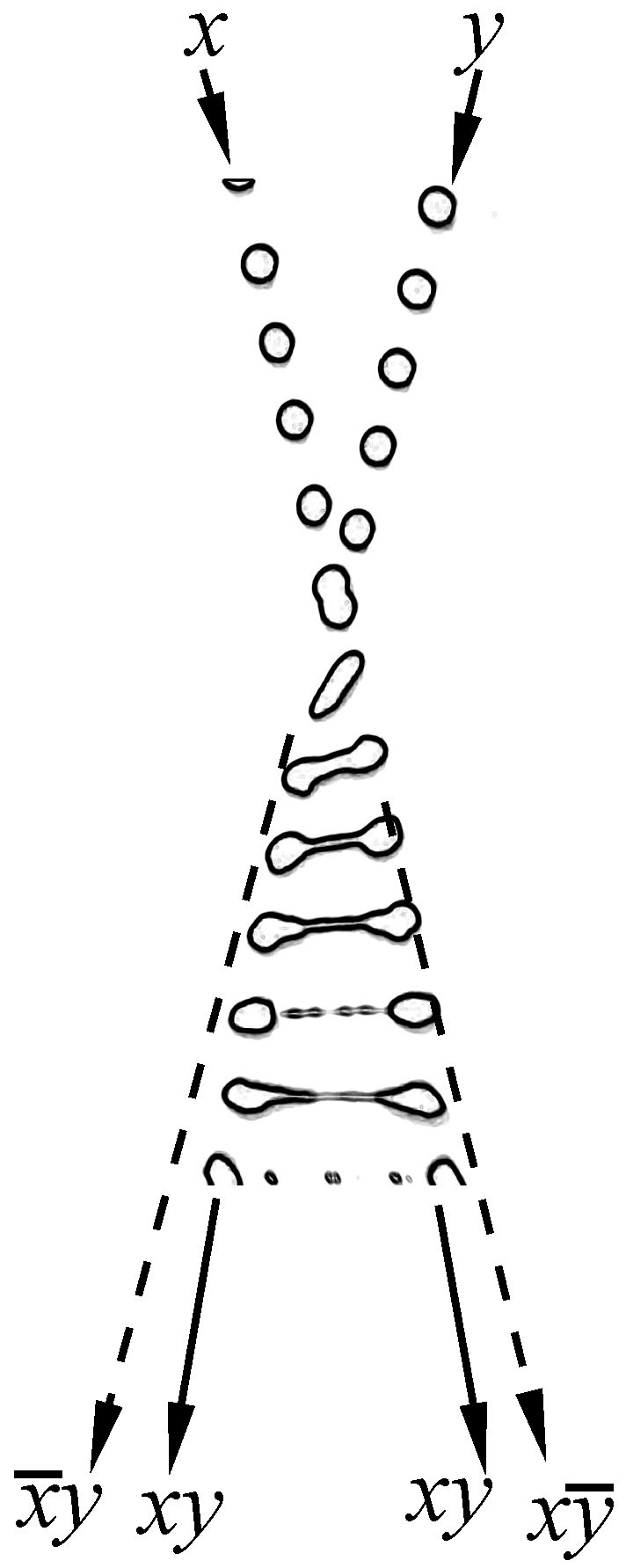}\label{FreeDropletsBinary_A}}
\subfigure[]{\includegraphics[scale=0.22]{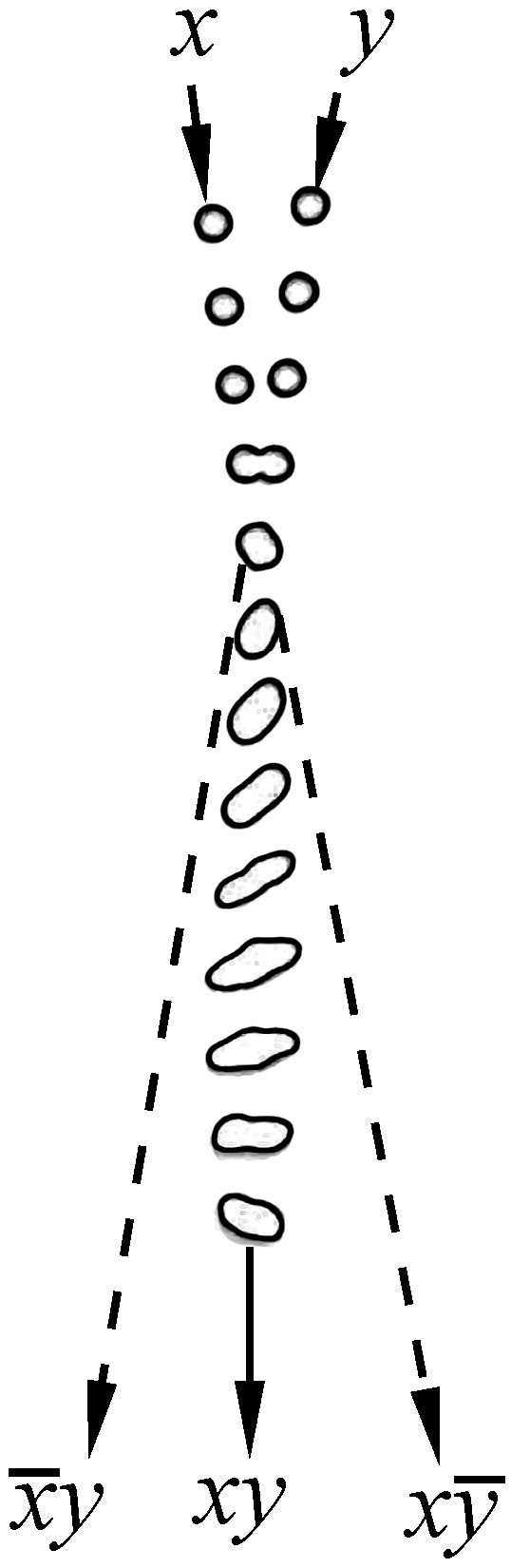}\label{FreeDropletsBinary_B}}
\subfigure[]{\includegraphics[scale=0.22]{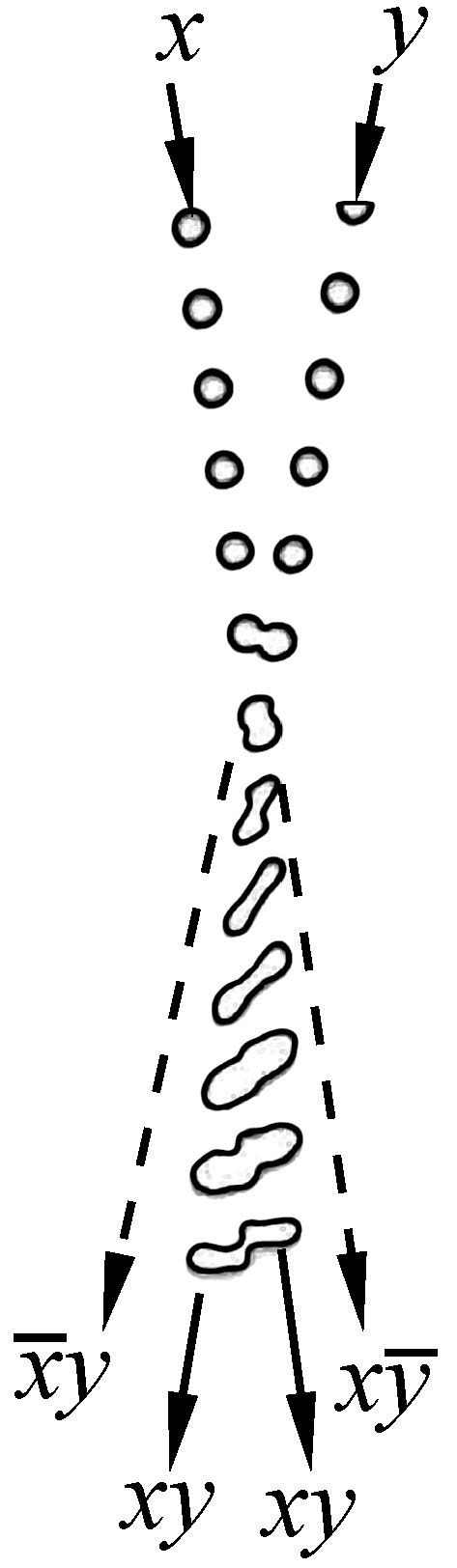}\label{FreeDropletsBinary_C}}
\subfigure[]{\includegraphics[scale=0.1]{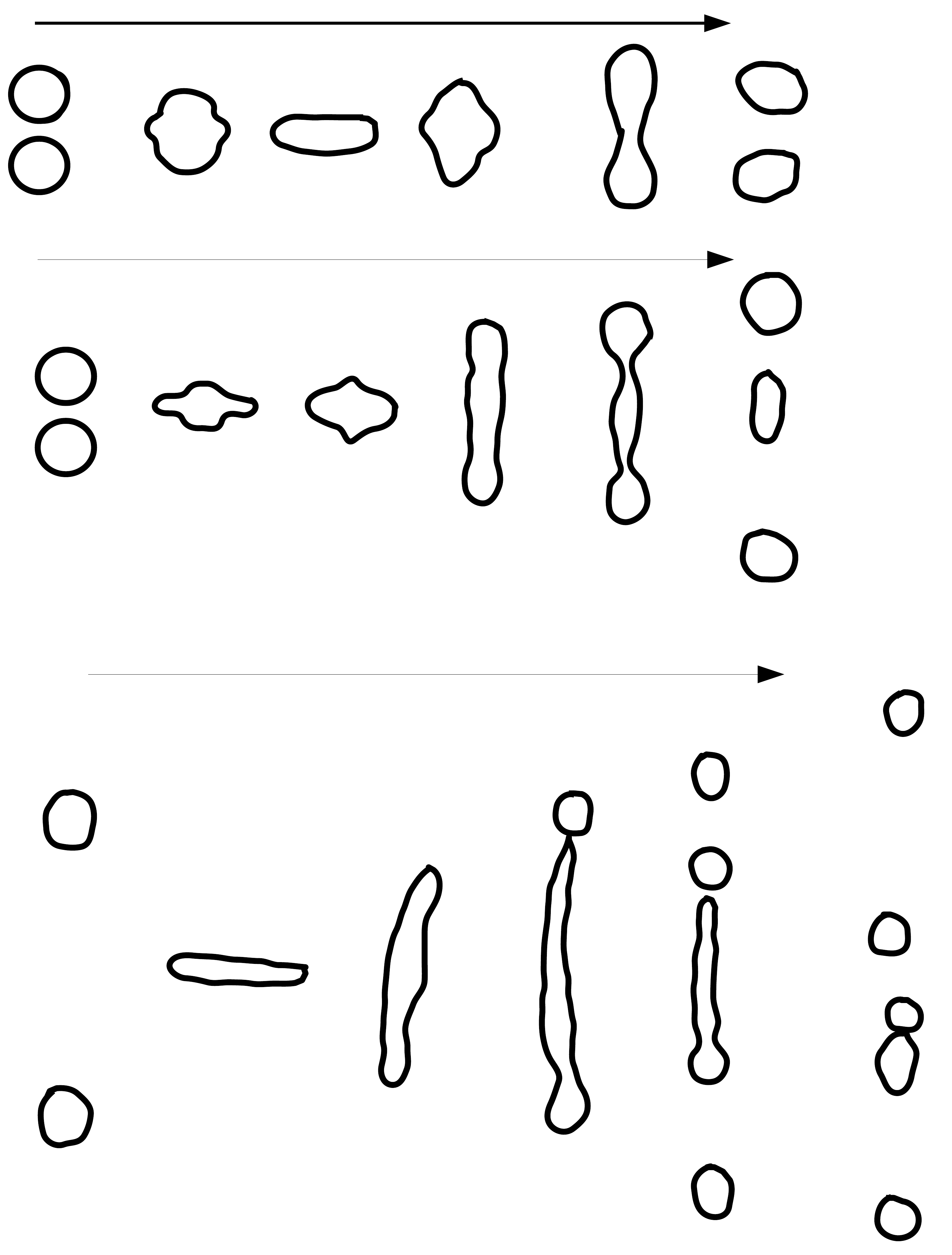}\label{ashgriz_droplets_A}}
    \subfigure[]{\includegraphics[scale=0.2]{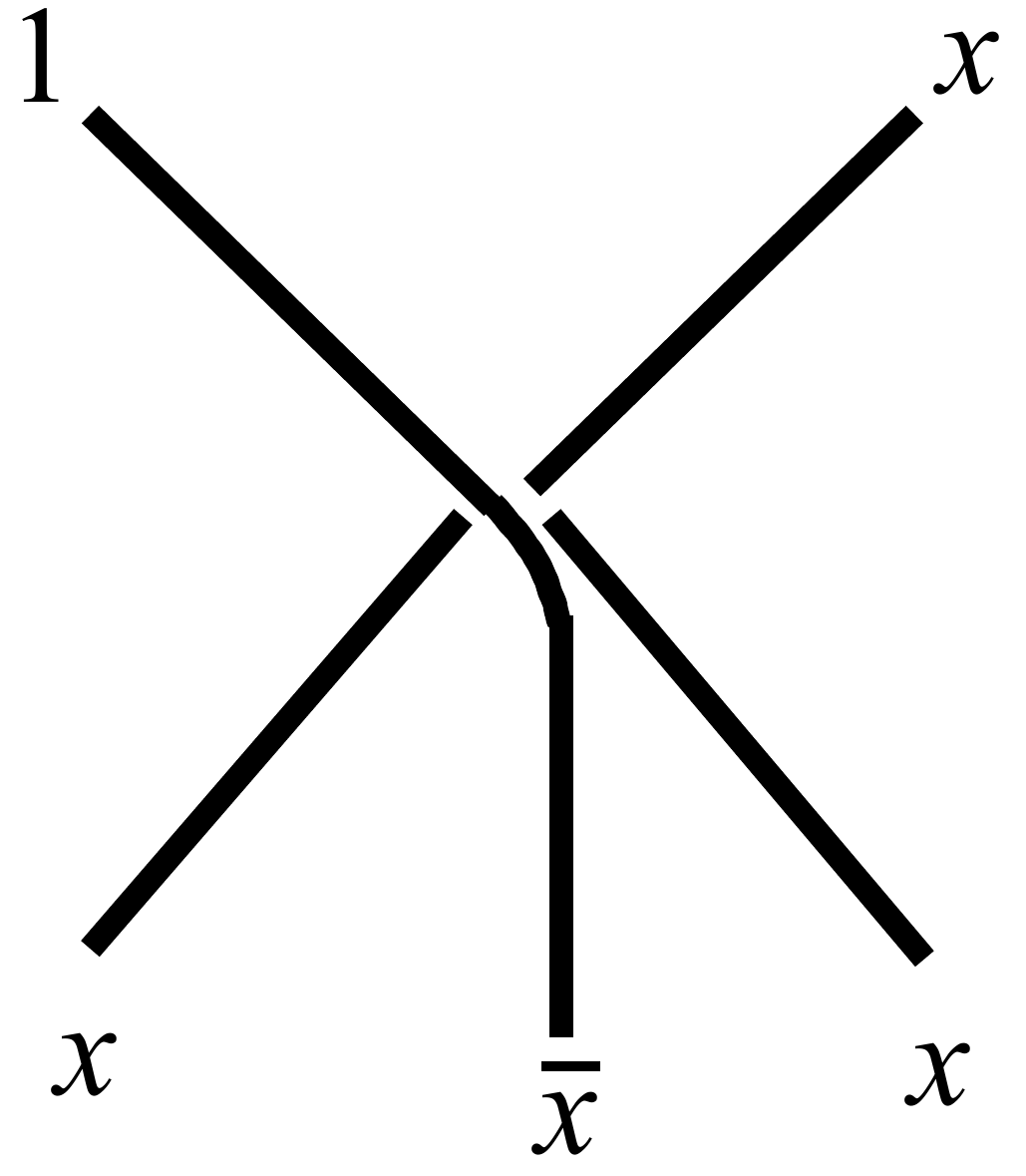}\label{DropletGate_NOT}}
    \subfigure[]{\includegraphics[scale=0.3]{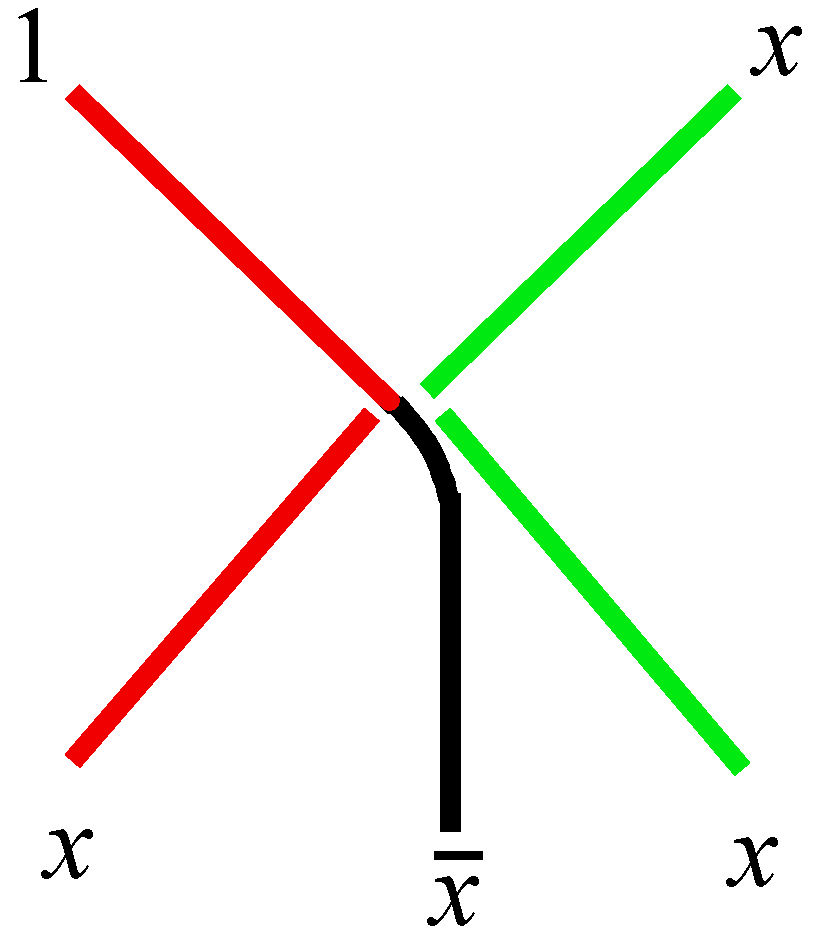}\label{Droplet_X}}
    \subfigure[]{\includegraphics[scale=0.3]{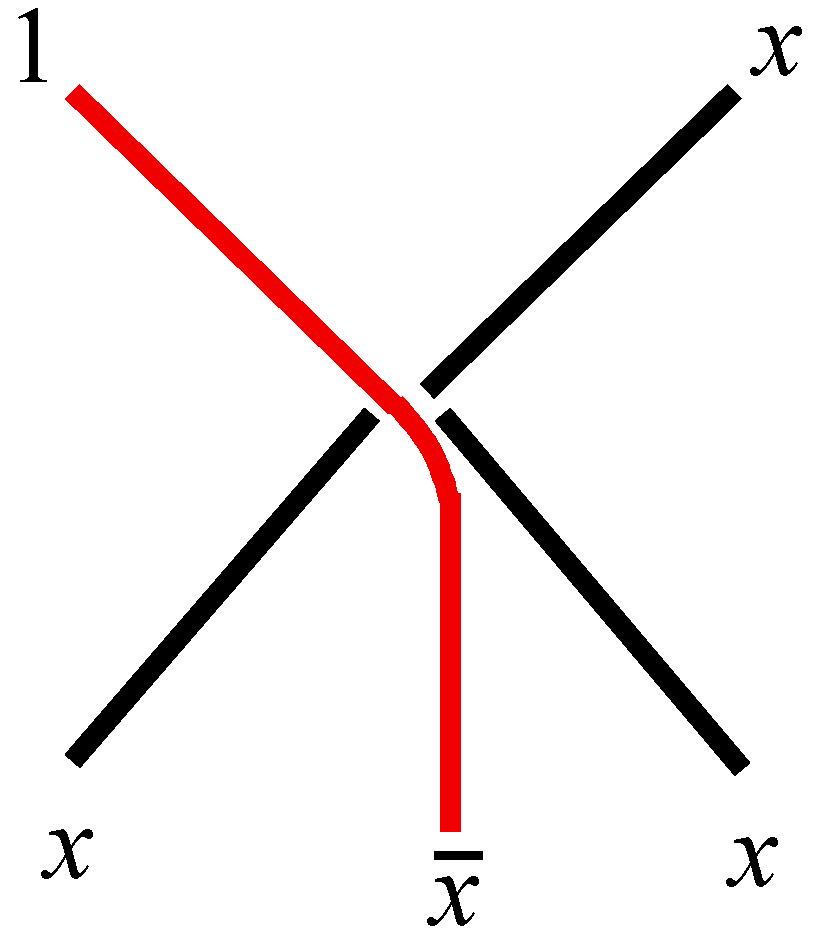}\label{Droplet_NOT_X}}
    \subfigure[]{\includegraphics[scale=0.2]{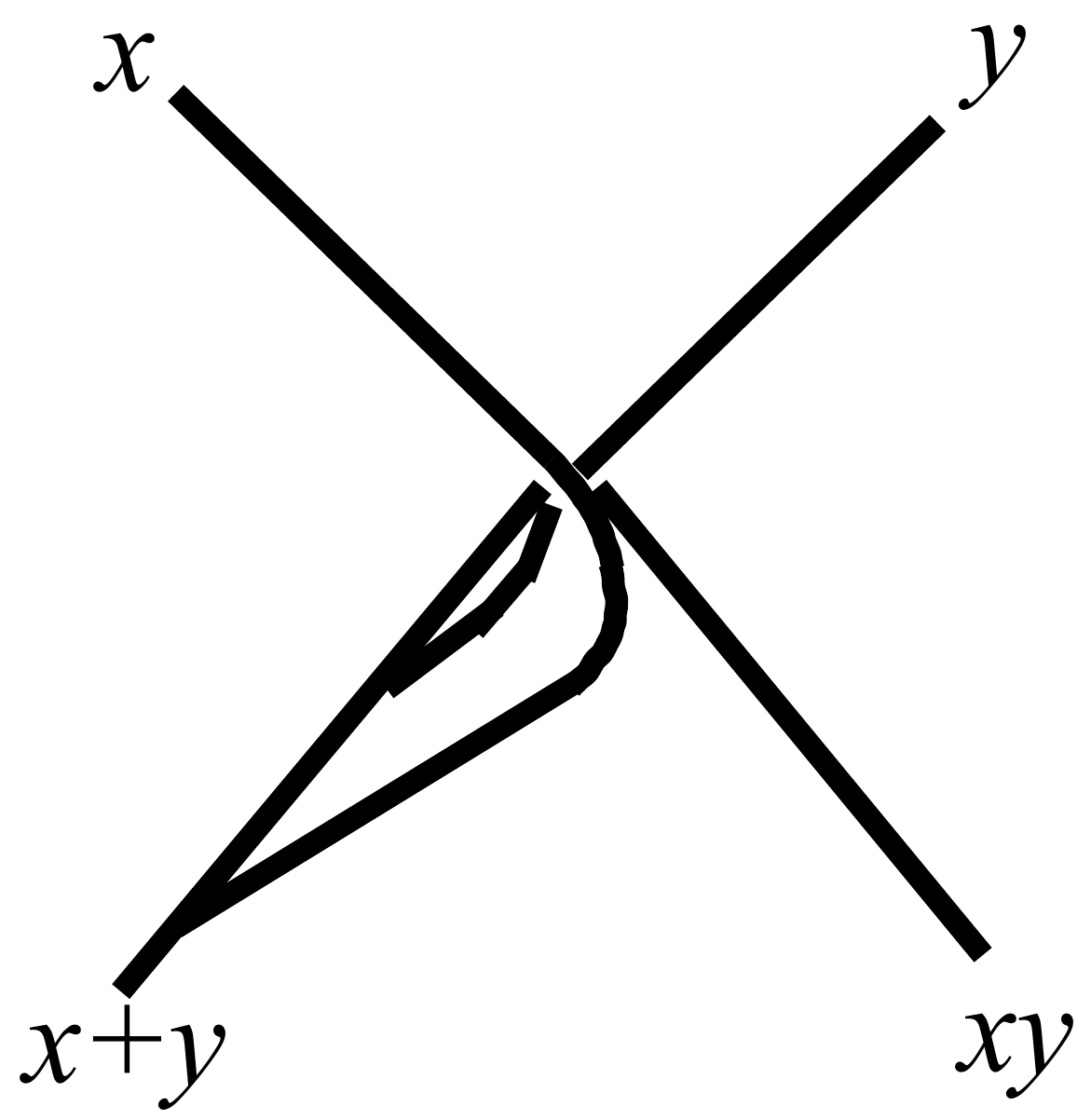}\label{DropletGate_OR_AND_scheme}}
    \subfigure[]{\includegraphics[scale=0.2]{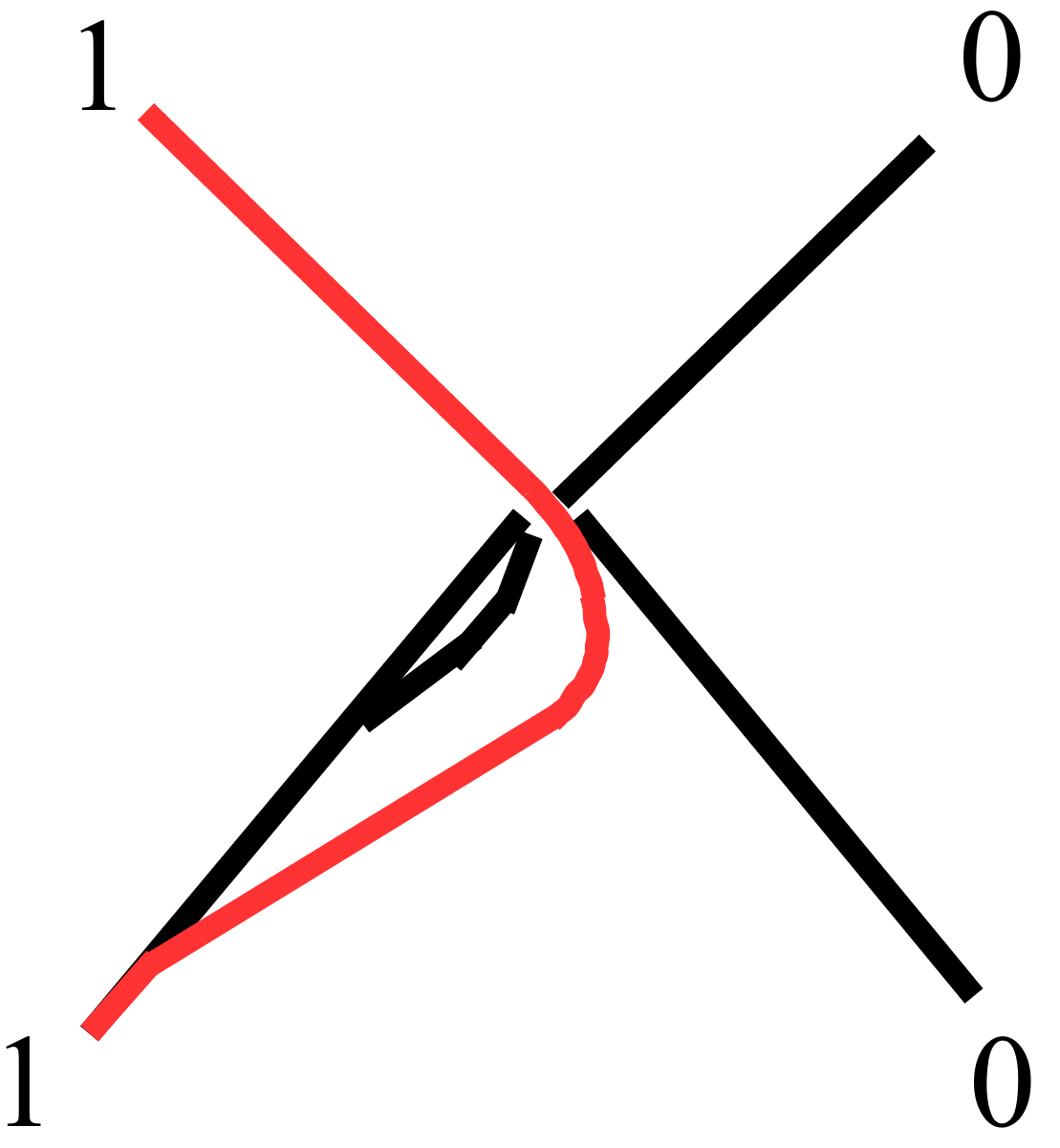}\label{DropletGate_OR_AND_10}}
    \subfigure[]{\includegraphics[scale=0.2]{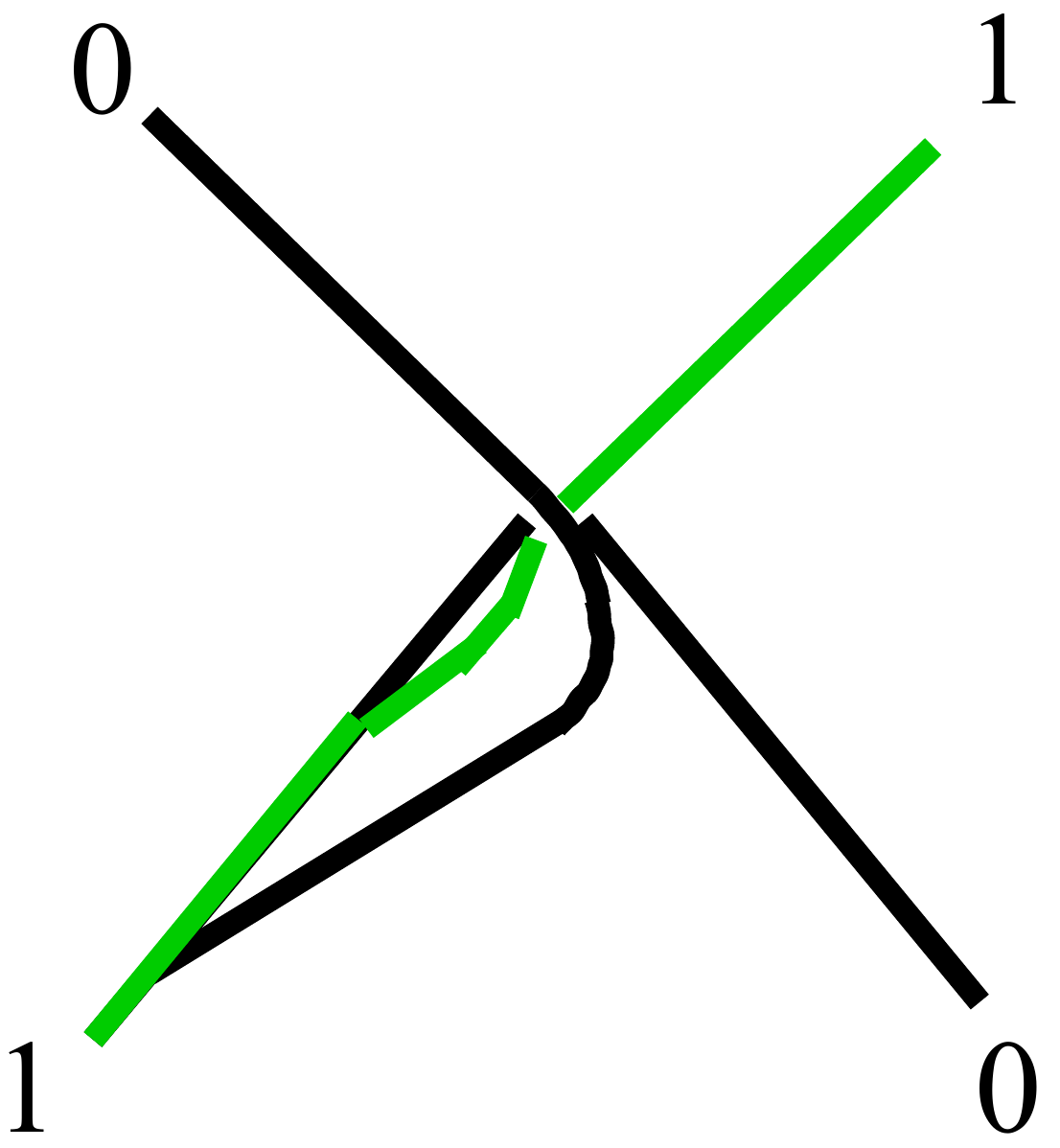}\label{DropletGate_OR_AND_01}}
    \subfigure[]{\includegraphics[scale=0.2]{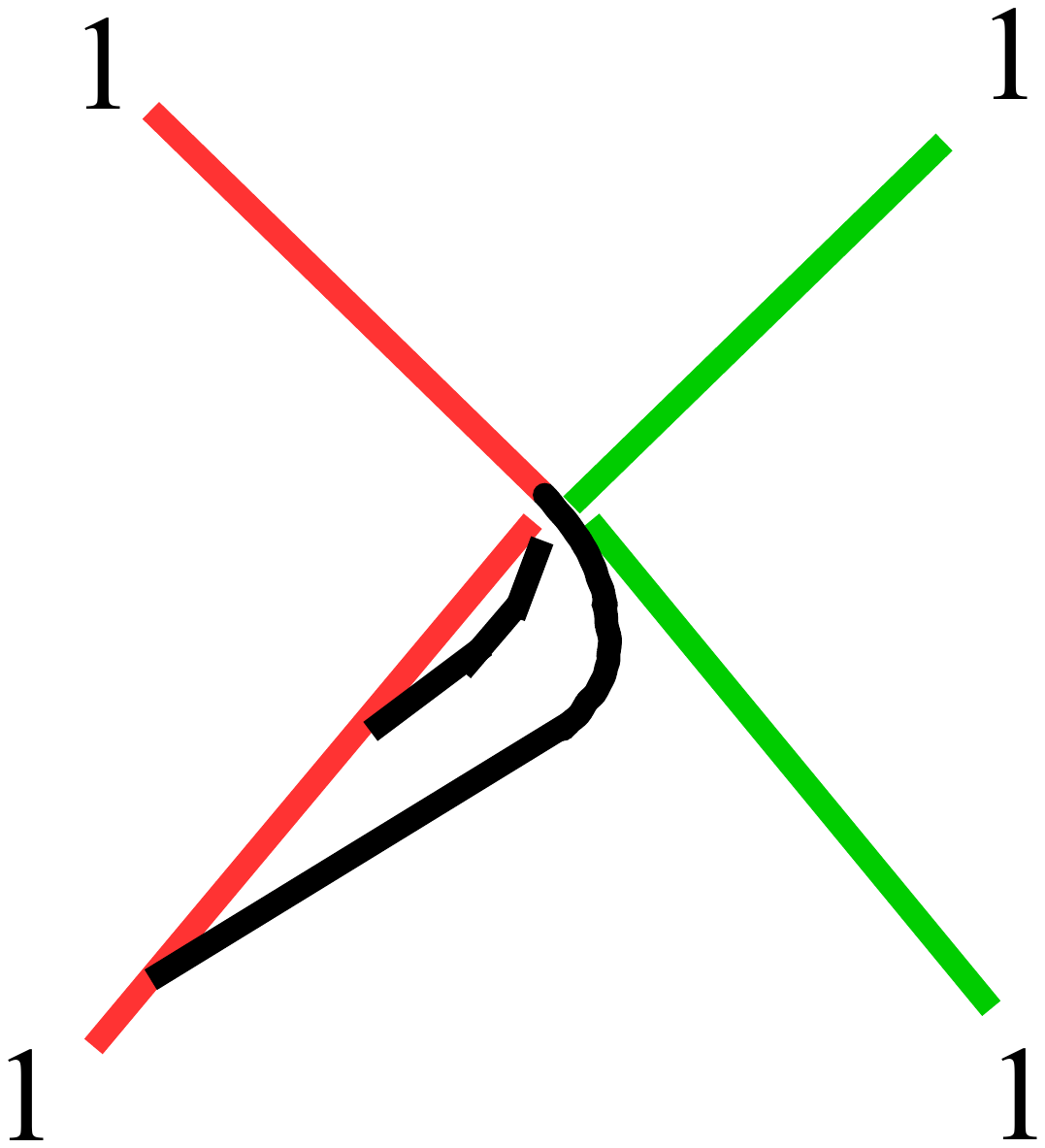}\label{DropletGate_OR_AND_11}}
\caption{Collision-based computing with droplets and liquid marbles. (a)~Hard balls gate, Fredkin gate. (b)~Soft balls gate, Margolus gate. (c)~Fusion/coalescence gate. (def)~Gates derived from laboratory experiments~\cite{rabe2010experimental} with droplets colliding with various offsets: (d)~stretching separation, (e)~coalescence, and (f)~reflexive separation. Illustrations of colliding droplets are redrawn from figure 4 of the paper ~\cite{rabe2010experimental}.(g)~Outcome of binary droplet collision experimentally found by Ashgriz and Poo \cite{ashgriz1990coalescence}; Weber numbers, from top to bottom are $We=23$, $We=40$ and $We=96$. Collisions are redrawn from \cite{ashgriz1990coalescence}.
(h--n)~Schemes of gates implemented with droplets on a hydrophobic surface in~\cite{mertaniemi2012rebounding}. (h--j)~{\textsc not} gate, scheme (h), $x=1$ (i), $x=0$ (j) (k--n)~{\sc or}-{\sc and} gate: scheme (k), $x=1, y=0$ (l), $x=0, y=1$ (m), $x=1, y=1$ (n); pathway of a droplet originated in channel $x$ is coloured red, and in channel $y$ green.}
\label{margolusgate}
\end{figure}

When two hard balls collide they reflect in such a manner that a trajectory of a reflected ball is at the angle less than 180\textsuperscript{o} to the trajectories of the same ball not involved in the collision (Fig.~\ref{HardBallsGate}). If presence of balls represent {\sc True} values of Boolean variables $x$ and $y$ then trajectories of the reflected balls represent $\overline{x}y$ and $x\overline{y}$. This was an inspiration for the Fredkin gate~\cite{fredkin2002conservative}. When balls are soft they compress on impact and propagate for some period of time as conjunct bodies. Then they restore their shapes and bounce back, thus their output trajectories are shifted in time-space (Fig.~\ref{SoftBallsGate}). When soft bodies impact into each at particular regimes they might merge into a single ball and lose their momentum (Fig.~\ref{MergingDropletsGate}): in this case we have only three output trajectories instead of four. The gate shown in Fig.~\ref{MergingDropletsGate} is a fusion gate analogous to Fig.~\ref{AND412}.  

Outcomes of the collisions between liquid droplets depend on the Weber number of the colliding droplets and their offset.  Binary collisions between droplets have been analysed exhaustively by Ashgriz and Poo in 1990~\cite{ashgriz1990coalescence} an twenty years later by Rabi et al.~\cite{rabe2010experimental}. Based on their results we can derive the following experimental droplet gates: stretching separation  (Fig.~\ref{FreeDropletsBinary_A}) and reflexive separation (Fig.~\ref{FreeDropletsBinary_A}) of colliding droplets represent soft balls gate, coalescence of droplets (Fig.~\ref{FreeDropletsBinary_B}) is a fusion gate (Fig.~\ref{MergingDropletsGate} and Fig.~\ref{AND412}). For certain parameters of the droplet collisions, one or more stationary droplets are formed~\cite{ashgriz1990coalescence} (Fig.~\ref{ashgriz_droplets_A}), they could represent results of the {\sc and} gate and even used as elementary memory units (presence of a stationary droplet is a bit up, absence is a bit down). 

Mertaniemi et al.~\cite{mertaniemi2012rebounding} explicitly interpreted collisions between droplets on a hydrophobic surface in terms of Boolean logic gates. They programmed trajectories of droplets, and thus architectures of the logical gates, by the geometry of grooves, on the hydrophobic surface, along which the droplets travel. Examples of two gates, the {\sc not} gate (Fig.~\ref{DropletGate_NOT}--\ref{Droplet_NOT_X}) and the {\sc or-and} gate (Fig.~\ref{DropletGate_OR_AND_scheme}--\ref{DropletGate_OR_AND_11}). In the {\sc not} gate (Fig.~\ref{DropletGate_NOT}) there is a droplet travelling along channel labelled `1'. When this droplet `1' collides with the droplet travelling along the channel $x$, the droplet `1' reflects into south-west channel $x$ (Fig.Z\ref{Droplet_NOT_X}. When input $x$ is {\sc False} and only droplet in channel `1' is present, the droplet `1' travels into the channel $\overline{x}$. The {\sc or-and} gate (Fig.~\ref{DropletGate_OR_AND_scheme}) shows several channels leaving the collision site thus that the droplets $x$ or $y$ are always routed to the channel $x+y$ when each of the droplets enters the gate alone (Fig.~\ref{DropletGate_OR_AND_10} and \ref{DropletGate_OR_AND_01}). When both droplets enter the gate, they collide. Then one droplet enters the channel $xy$  while other droplet still travels to the channel $x+y$ (Fig.~\ref{DropletGate_OR_AND_11}).

\begin{figure}[!tbp]
\centering
\subfigure[]{\includegraphics[height=2.2cm]{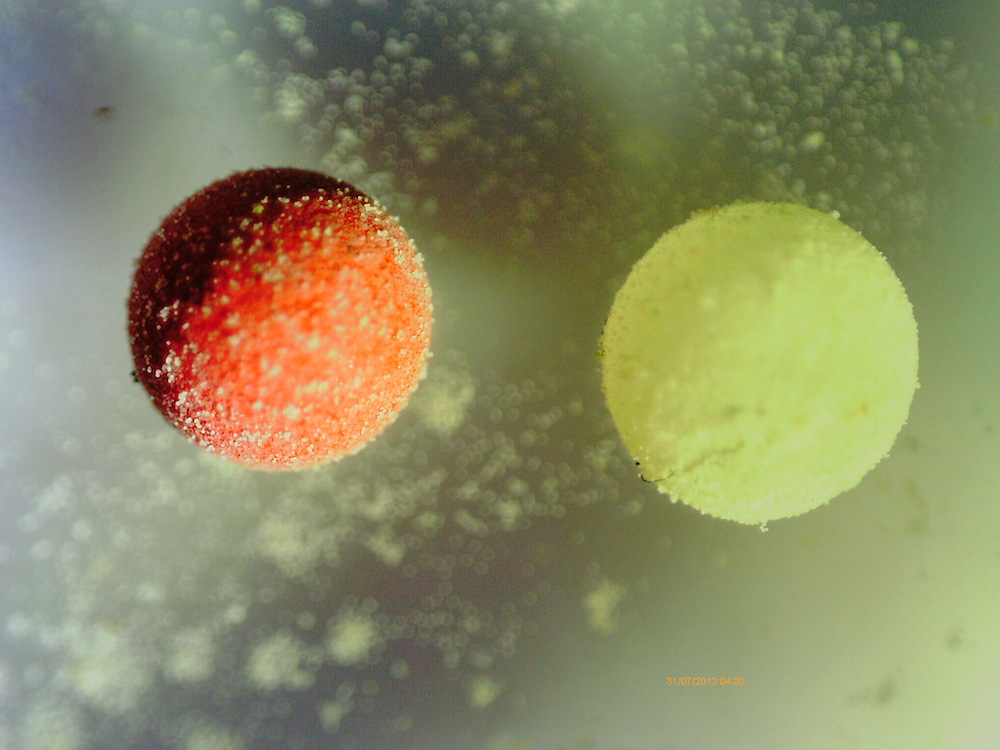}\label{redyellowmarble}}
\subfigure[]{\includegraphics[height=2.7cm]{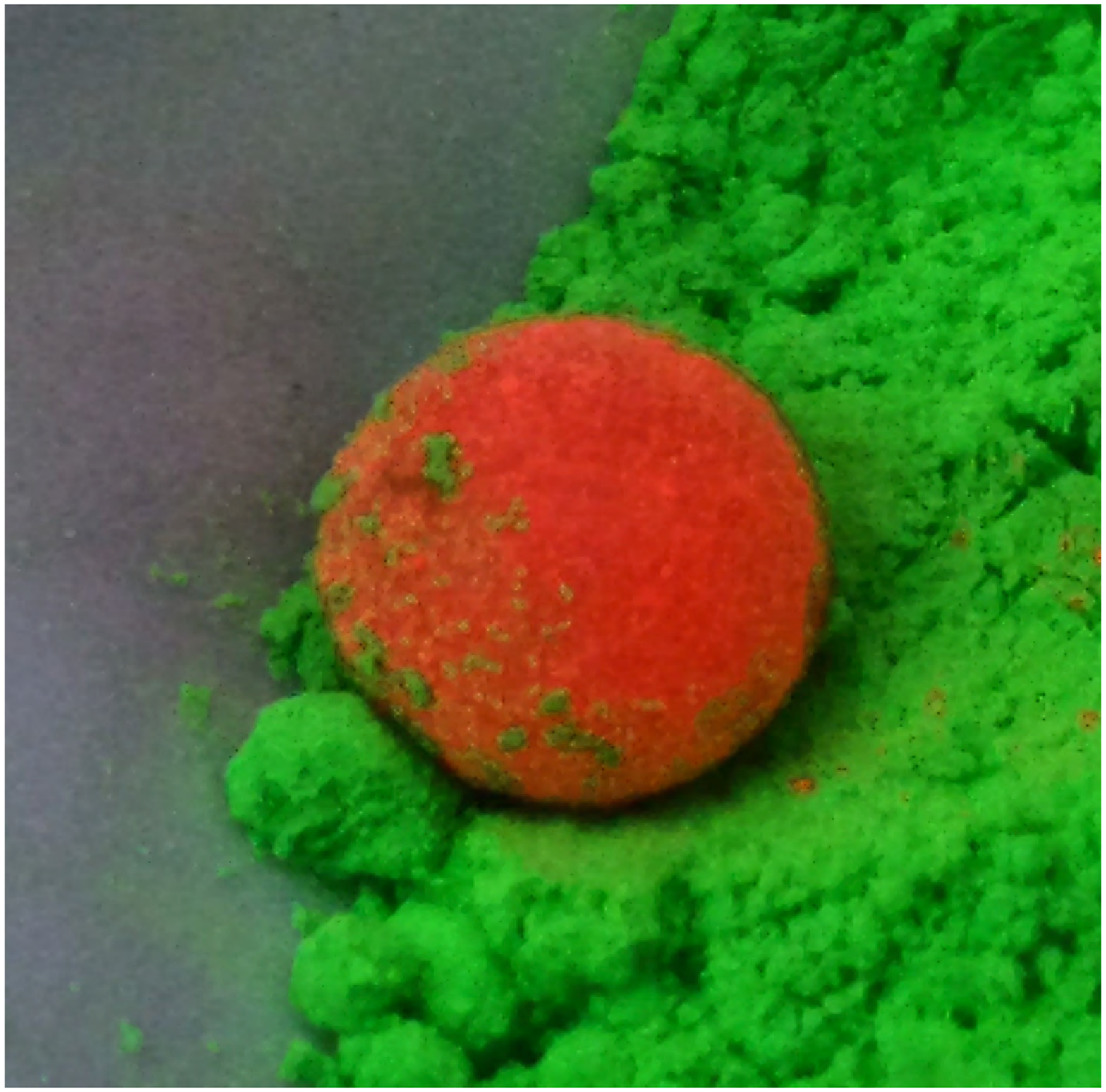}\label{RedMarble}}
\subfigure[]{\includegraphics[height=2.7cm]{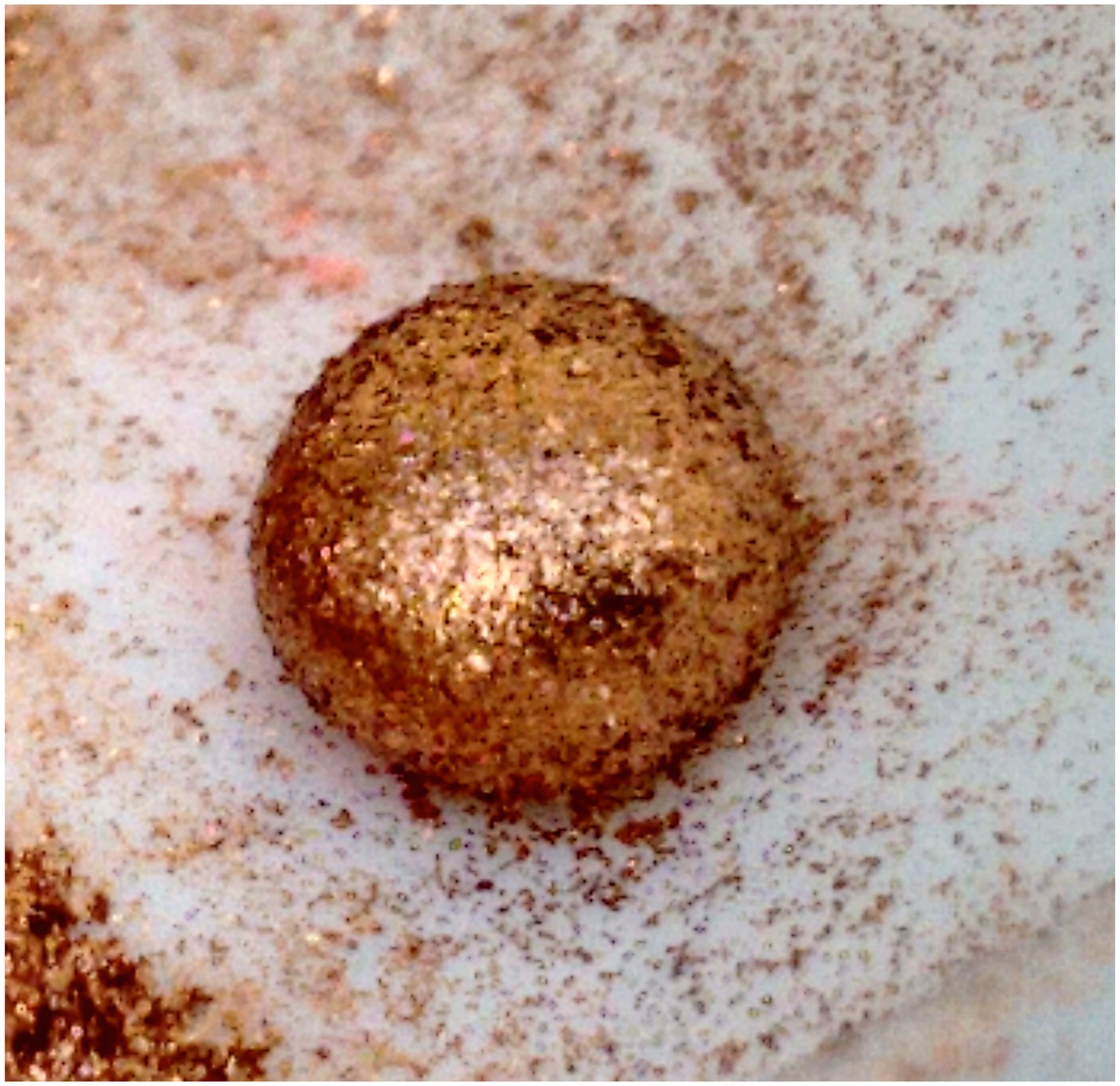}\label{CopperMarble}}
\subfigure[]{\includegraphics[width=0.24\textwidth]{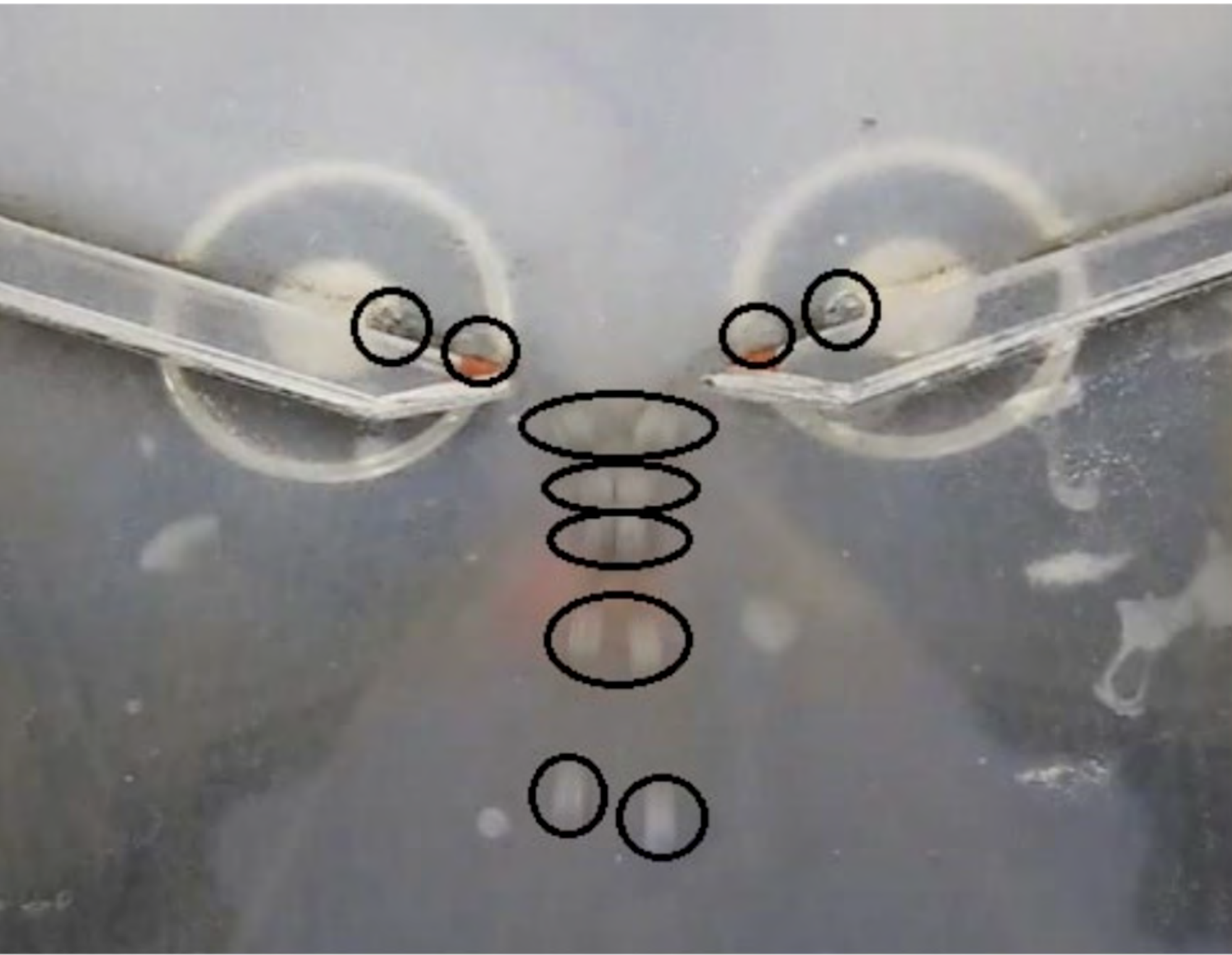}\label{tomgate}}
\subfigure[]{\includegraphics[width=0.2\textwidth]{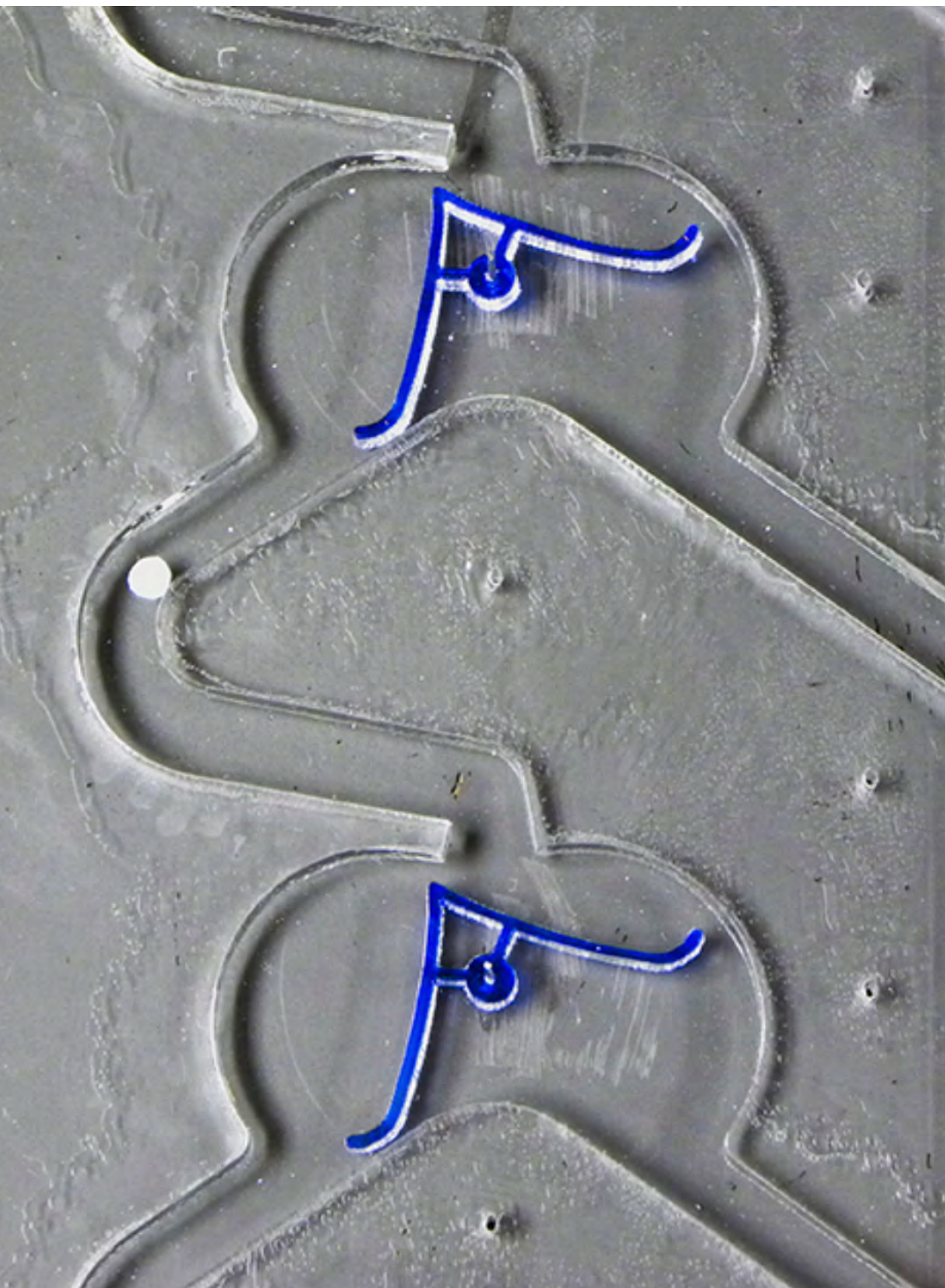}\label{tomcounter}}
\caption{(a)~Liquid marbles with Lycopodium coating and water cargo on right and a mix of red Indian ink and water on the left;  marbles are c. 5 $\mu$l in volume.
(b)~50~$\mu$l water marble coated by red fluorescent fingerprint powder. 
(c)~50~$\mu$l water marble coated by copper particles.
(c)~Soft balls gate, Fig.~\ref{SoftBallsGate}, implemented with liquid marbles, filmed by a high-speed camera; the gate is prototyped by Draper et el.~\cite{draper2017liquid}. From~\cite{draper2017liquid}. (d)~A fragment of the liquid marble binary counter presented in~\cite{draper2018mechanical}.
}
\label{marbles}
\end{figure}

Instead of running droplets on a hydrophobic surface we can coat the droplets with a hydrophobic power. The droplets then become liquid marbles. The liquid marbles, proposed by  Aussillous and Qu\'{e}r\'{e} in 2001~\cite{aussillous2001liquid}, are liquid droplets coated by hydrophobic particles at the liquid/air interface (Fig.~\ref{redyellowmarble}--\ref{CopperMarble}). In 2016 Adamatzky proposed to make experimental laboratory prototypes of computing devices allowing the liquid marbles to explore additional degrees of freedom to travel in different directions~\cite{EPSRCmarbles}. First collision based logical gate with liquid marbles was prototypes by Draper et al.~\cite{draper2017liquid}. An example of the liquid marble gate in action is shown in Fig.~\ref{tomgate}. 

Computing schemes involving `proper' collision-based gates require a synchronisation of signals. Such synchronisation is indeed achievable but place additional burden on preparation of data for computation; thus, simultaneously with developments of synchronous circuits  we considered producing experimental prototypes of asynchronous devices based on cantilever moving parts, actuated by liquid marbles~\cite{EPSRCmarbles}.  

In 1965 J.~T.~Goodfrey~\cite{godfrey1968binary} proposed a mechanical binary digital computer: the configuration of flip-flops arranged on an inclined surface and operated by metal balls rolling down the surface.  This design inspired us to prototype an asynchronous binary counter actuated by liquid marbles~\cite{draper2018mechanical} (Fig.~\ref{tomcounter}). 

Droplets and liquid marbles  devices use gravity force to operate. This can be avoided by using self-propulsive liquid marbles, e.g. aqueous ethanol marble on a water surface~\cite{bormashenko2015self,ooi2015floating} or polypyrrole and  carbon black marbles driven by light~\cite{paven2016light}.
%\enlargethispage{20pt}

\section{Reaction-diffusion computers}
\label{rectiondiffusioncomputers}

A reaction-diffusion computer~\cite{adamatzky2005reaction,adamatzky2011topics,adamatzky2012reaction} is a spatially extended chemical system which processes information
by transforming an input concentration profile to an output concentration profile in a deterministic and controlled  manner. In reaction-diffusion computers the data are represented by concentration profiles of reagents,  information is transferred by propagating diffusive and phase waves, computation is implemented via the interaction 
of these travelling patterns (diffusive and excitation waves), and results of the computation are recorded as a final concentration profile. Chemical reaction-diffusion computing is amongst the leaders in providing experimental prototypes  in the fields of unconventional and nature-inspired computing.

\begin{figure}[!tbp]
    \centering
    \subfigure[]{\includegraphics[width=0.3\textwidth]{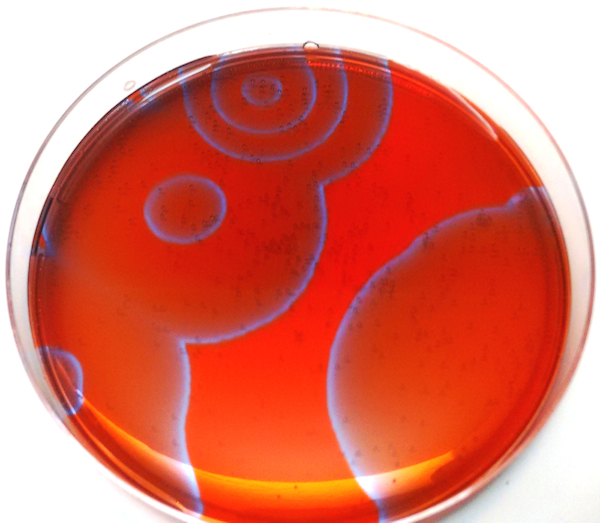}\label{bzdish}}
    \subfigure[]{\includegraphics[angle=90,width=0.3\textwidth]{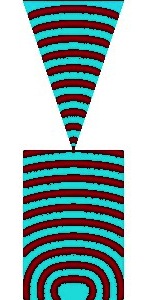}\label{diodeforward}}
    \subfigure[]{\includegraphics[angle=90,width=0.3\textwidth]{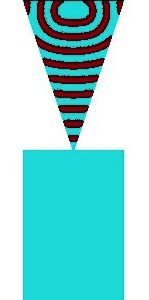}\label{diodebackward}}
    \subfigure[]{\includegraphics[width=0.45\textwidth]{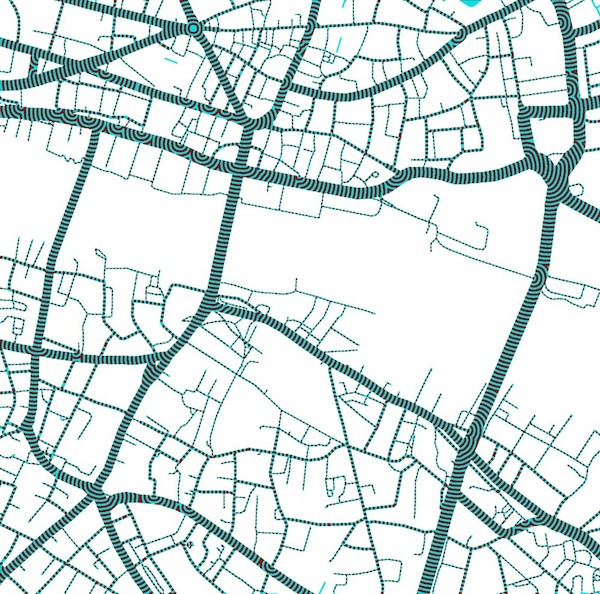}\label{londonexcitable}}
        \subfigure[]{\includegraphics[width=0.45\textwidth]{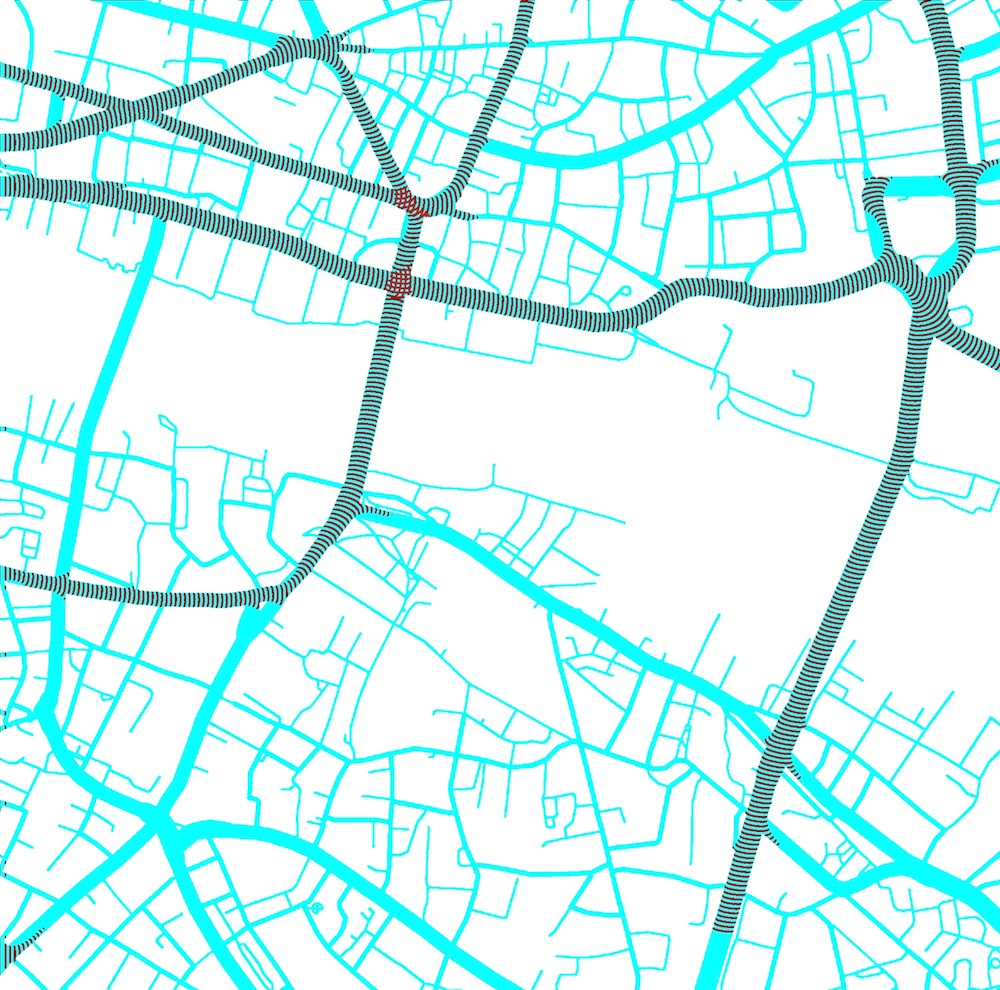}\label{londonsubexcitable}}
          \subfigure[]{\includegraphics[width=0.27\textwidth]{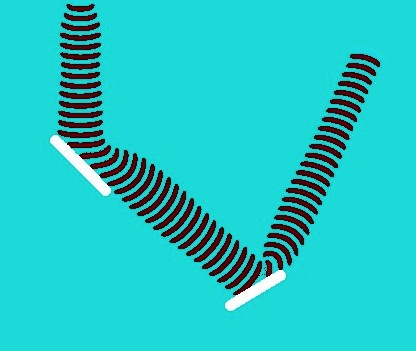}\label{reflection3}}
          \subfigure[]{\includegraphics[width=0.27\textwidth]{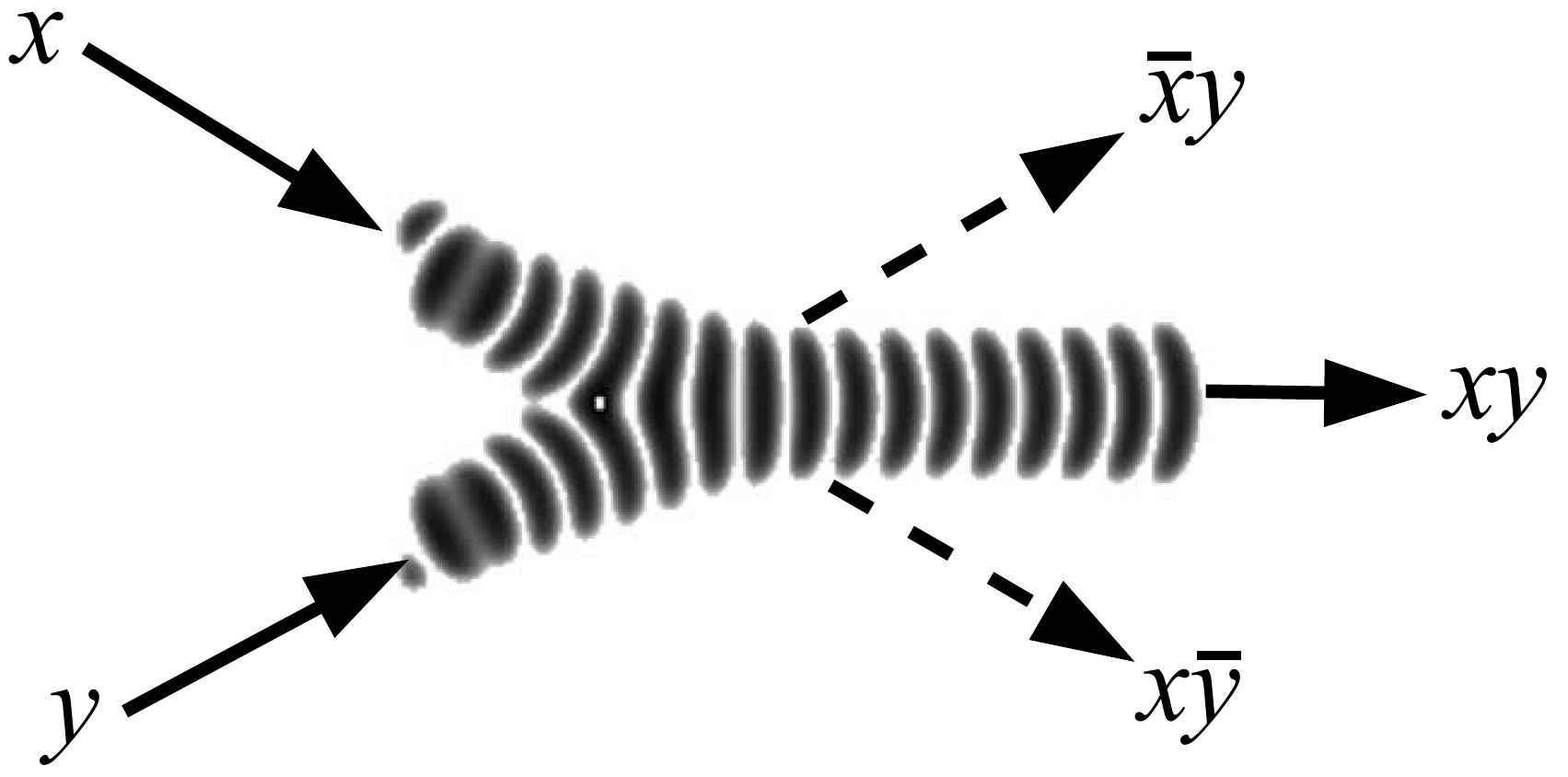}\label{CollisionGate}}
        \subfigure[]{\includegraphics[width=0.3\textwidth]{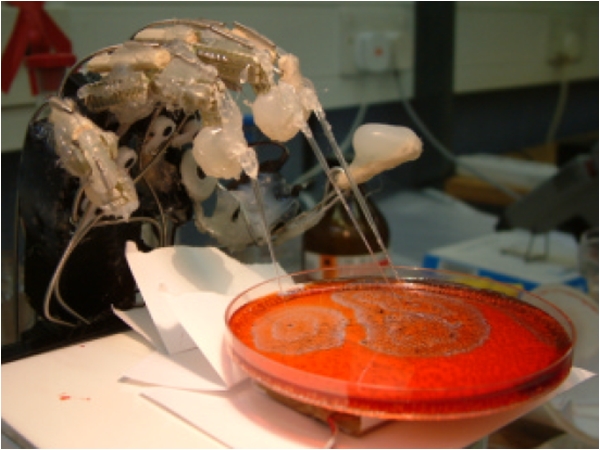}\label{BZhand}}
    \caption{(a)~A thin layer BZ reaction, travelling oxidation wave-fronts are blue. (bc)~Computer model of Agladze diode. (b)~Forward propagation. Excitation wave-front propagates from the right to the left. (c)~Backward propagation. Excitation wave-front propagates from the let to the right. (de)~Exploration of London street networks with fully excitable (d) and sub-excitable medium (e). Time lapse snapshots two-variable Oregonator model. 
     (f)~Routing of a wave-fragment in a sub-excitable medium. The medium is sub-excitable but two rectangular shapes (white rectangles) are not excitable. 
Grid size is 500$\times$500 nodes. 
The pictures are not snapshots of many wave fronts generated at the initial stimulation point but time lapsed snapshots of a single wave-fragment  recorded every $150^{th}$ step of numerical integration. 
    (g)~Collision-based gate. Case shown when both inputs have value {\sc True} thus two wave-fragments collide. Time lapse overlays. The wave-fragments collide and fuse into a new localised excitation travelling east. 
    (h)~Robotic hand, whose glass capillary tubes-nails release small quantities of colloidal silver, interacts with Belousov-Zhabotinsky medium.   From~\cite{yokoi2004excitable}.}
    \label{fig:bz}
\end{figure}

Belousov-Zhabotinsky (BZ) reaction, a periodical oxidation of malonic acid in solution~\cite{belousov1959,zhabotinsky1964periodical,zaikin1970concentration}, was a key substrate for implementing reaction-diffusion computers in last nearly fort years. In these computing devices an information is represented by travelling oxidation wave fronts (Fig.~\ref{bzdish}) and the computation is programmed by geometrical constraining of the medium or configurations of the excitation initiation sites. 

Research on BZ based information  processing has been started in mid-1980s when Kuhnert, Krinsky and Agladze demonstrated that a thing-layer of light-sensitive BZ reaction can implement contrast modification, detection of a contour and smoothing of the half-tone images projected onto the medium~\cite{kuhnert1986new,kuhnert1989image}. These results ignited an almost forty years epoch of information processing with BZ medium. Works on image processing with BZ continued till early 2000s~\cite{aliev1994oscillation,agladze1995phase,rambidi1998information,rambidi2002image,rambidi2005biologically}. 

In mid-1990s Showalter laboratory produced a series of experimental prototypes of logical gates implemented via interaction of oxidation wave-fronts in geometrically constrained BZ medium~\cite{steinbock1996chemical}. Numerous, modelled and/or implemented in experimental laboratory, devices followed, including signal switches~\cite{sielewiesiuk2001logical}, counters~\cite{gorecki2003chemical}, one-bit adder~\cite{costello2011towards}, many-bit binary adder~\cite{zhang2012towards} and decoder~\cite{sun2013multi}, three-valued logic gates~\cite{motoike2005three},  square root approximation~\cite{stevens2012time}. 

Designs of potential computing circuits, was facilitated by a discovery of a chemical diode~\cite{agladze1996chemical}.
The diode was made of two plates covered with excitable solution.  The corner of the one plate was close to the plane side of 
another plate. Excitation wave-front  travelling in the forward direction reaches the contact site between the plates in  a state of a planar wave, it propagates through the contact site and then continues its expanding in the triangular part of the device (Fig.~\ref{diodeforward}). The wave-fragment travelling in the backward direction slows down while propagating towards the corner of the triangular plate (Fig.~\ref{diodebackward}). At the contact point size of the wave-fragment becomes so small, for the level of medium's excitability, that it annihilates without crossing the contact site between the plates.

The above prototypes constrained BZ medium in templates, that is BZ computers were programmed by architecture. There is another option to realise logical gates in BZ. This is a  collision-based, or dynamical, computation~\cite{fredkin2002conservative,adamatzky2002collision}, see Sect.~\ref{liquidmarbles}.  In 2001 Sendi\"{n}a-Nadal et al.\cite{sendina2001wave} experimentally demonstrated an  existence of localised excitations -- travelling wave fragments which behave like quasi-particles in the photosensitive sub-excitable BZ medium. In early 2000s we proposed that the compact wave-fragments can represent Boolean values and execute logical gates by colliding with each other~\cite{adamatzky2004collision}. The wave-fragments can be routed in the medium by using non-excitable domains as reflectors (Fig.~\ref{reflection3}). We employed the localisations to construct several logical gates and circuits~\cite{adamatzky2004collision,costello2005experimental,adamatzky2007binary,de2009implementation,toth2010simple}.

The BZ medium is also used for optimisation tasks, e.g. to assist a maze solving~\cite{steinbock1995navigating,agladze1997finding,rambidi2001chemical} and to calculate a shortest collision-free path~\cite{adamatzky2002collision}. Recently we found that by tuning excitability of the medium we can select critical features of street networks~\cite{adamatzky2018street,adamatzky2018exploring}. In excitable BZ, oxidation wave-fronts traverse all streets of the network (Fig.~\ref{londonexcitable}). A pruning strategy adopted by the medium with decreasing excitability when wider and ballistically appropriate streets are selected (Fig.~\ref{londonsubexcitable}).

In the early 2000s the first ever excitable chemical controller mounted on-board a wheeled robot was constructed and tested under experimental laboratory conditions~\cite{adamatzky2002collision,adamatzky2003liquid,adamatzky2004experimental},  and also a robotic hand was interfaced and controlled using a BZ medium~\cite{yokoi2004excitable} (Fig.~\ref{BZhand}). Litschel et al.~\cite{litschel2018engineering} shown that  by linking micro-reactor with BZ and establishing excitation and inhibitory connections between the neighbouring reactors, it is possible to generate travelling patterns of oscillatory activity resembling the neural locomotive patterns.

Control of a robot with oxidation wave-fronts can be embedded directly in the actuating materials. This can be done by impregnating a pH-sensitive gel with a catalyst and immersing it in the catalyst free BZ solution: a peristaltic motion will be observed~\cite{murase2008design}; such approach can be advanced to other types of smart materials and soft robots~\cite{yoshida2010self,balazs2014reconfigurable}.

\begin{figure}[!tbp]
\centering
\subfigure[]{\includegraphics[scale=0.25]{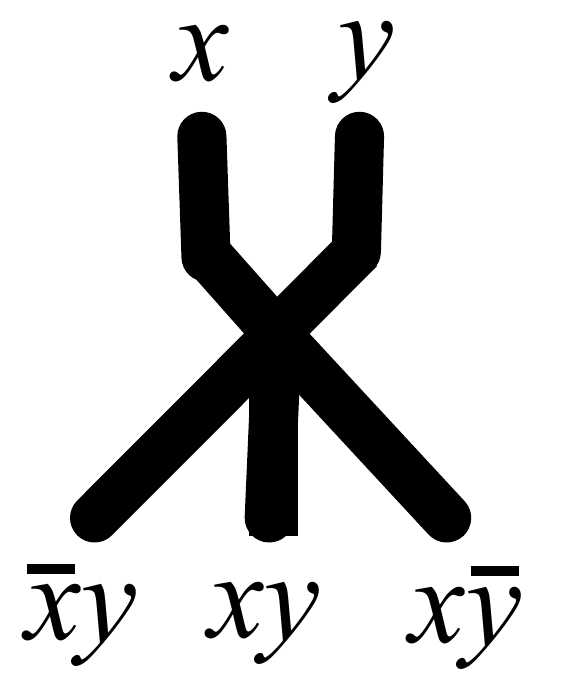}}
\subfigure[]{\includegraphics[scale=0.19]{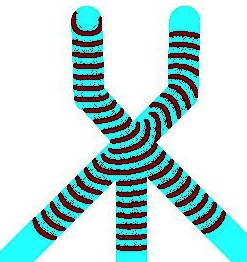}\label{excitable10}}
\subfigure[]{\includegraphics[scale=0.19]{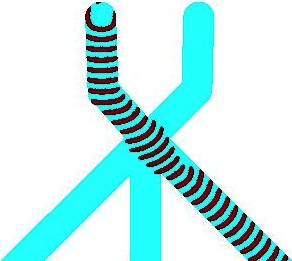}\label{subexcitable10}}
\subfigure[]{\includegraphics[scale=0.19]{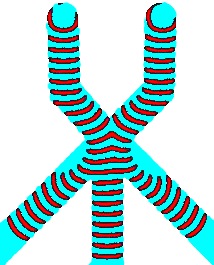}\label{excitable11}}
\subfigure[]{\includegraphics[scale=0.19]{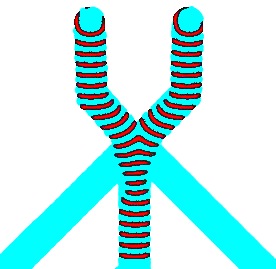}\label{subexcitable11}}\\
\subfigure[]{\includegraphics[scale=0.3]{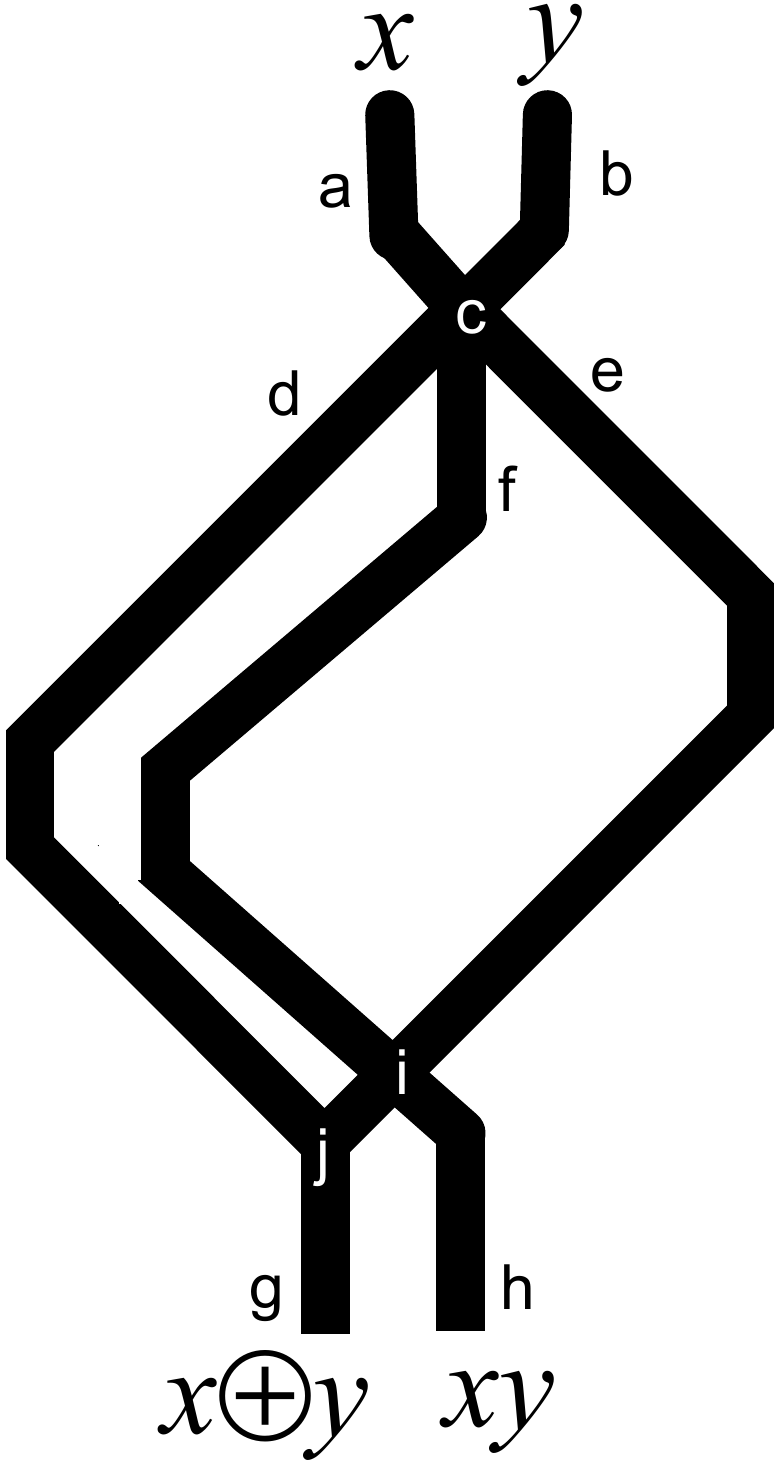}\label{bzhalfadder}}
\subfigure[]{\includegraphics[scale=0.15]{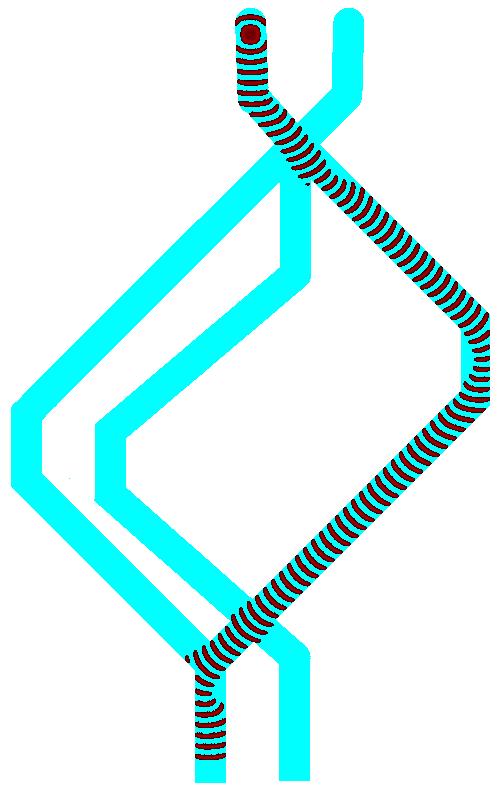}\label{bzhalfadder10}}
\subfigure[]{\includegraphics[scale=0.15]{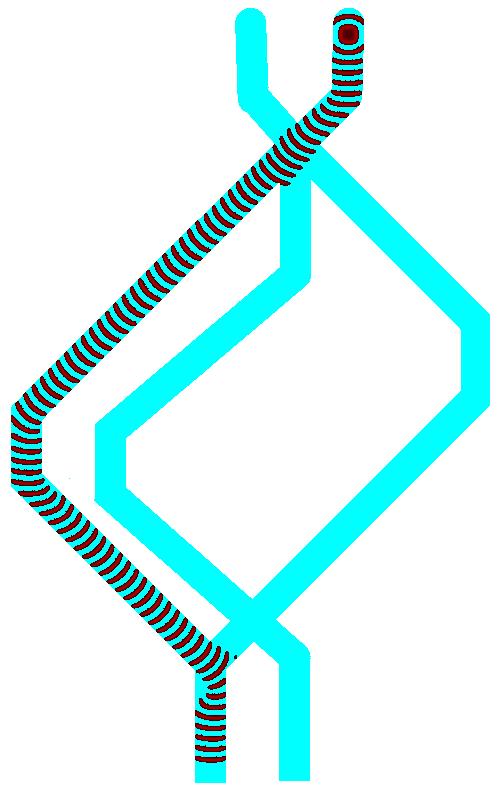}\label{bzhalfadder01}}
\subfigure[]{\includegraphics[scale=0.15]{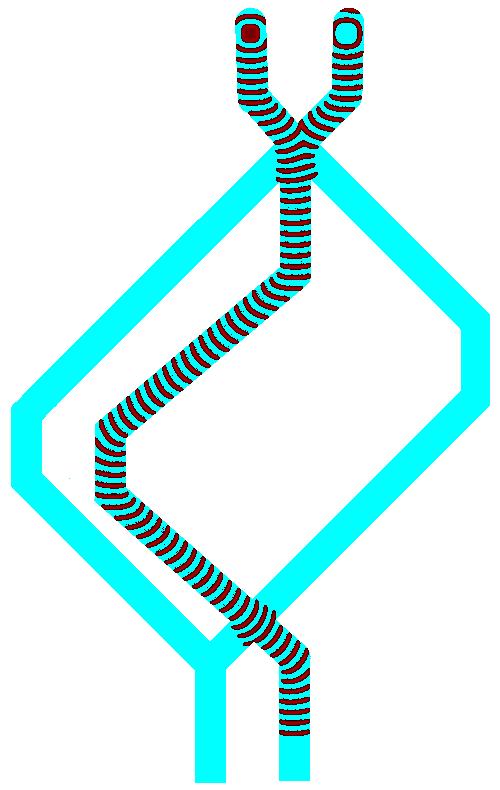}\label{bzhalfadder11}}
\subfigure[]{\includegraphics[scale=0.2]{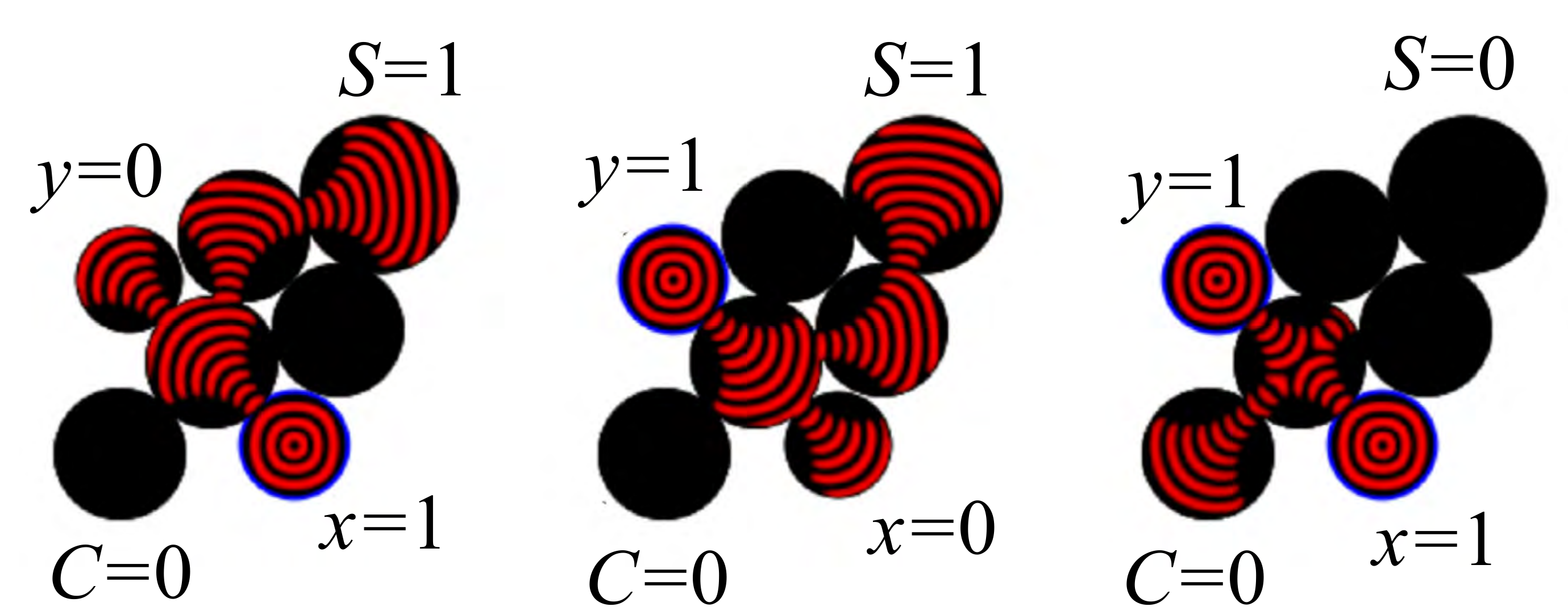}\label{BZDropletAdder}}
\subfigure[]{\includegraphics[scale=0.5]{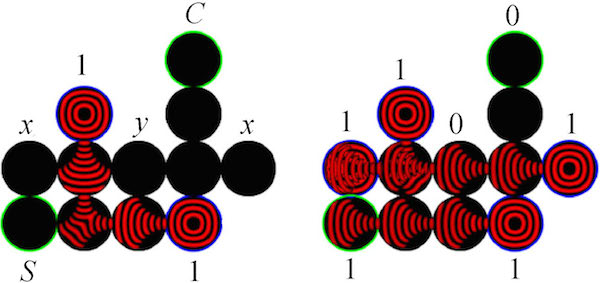}\includegraphics[scale=0.5]{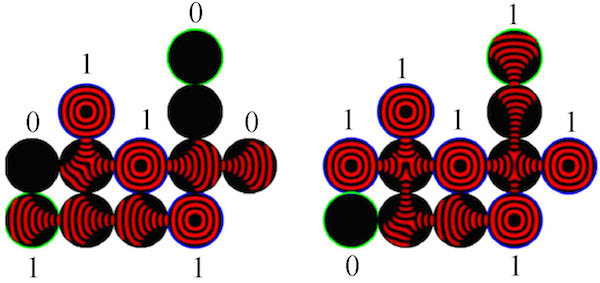}\label{BZDropletAdder_v2A}}
\caption{(a--e) Fusion gate. 
(a)~Scheme: inputs are $x$ and $y$, outputs are $xy$, $\overline{x}y$, $x\overline{y}$.  
(a--d)~Time lapse overlays of excitation waves.
(b)~Excitable mode, $x=1$, $y=0$.
(c~)~Sub-excitable mode, $x=1$, $y=0$.
(d~)~Excitable mode,  $x=1$, $y=1$.
(e)~Sub-excitable mode,  $x=1$, $y=1$.
(f)~A scheme of a one-bit half-adder: input channels are $a$ and $b$, output channels are $g$ and $h$, internal channels $d$, $e$, $f$, and junctions are $c$, $i$ and $j$. Input variables $x$ and $y$ are fed into channels $a$ and $b$, results $x \oplus y$ and $xy$ are read from channels $g$ and $h$.
(g--i) Time lapse overlays of excitation waves propagating in the one-bit half-adder for inputs (g)~$x=1$, $y=0$, (h)~$x=1$, $y=0$, (i)~$x=1$, $y=1$. 
 Sites of initial segment-wise perturbation are visible as discs. 
 Grid size is 500$\times$790 nodes.  See details in \cite{adamatzky2015binary}.
 (j) Composite half-adder implemented in 2D BZ vesicles: $S = x \oplus y, C = x \cdot y$, where inputs and outputs are all connected to a central reactor disc which can achieve both the {\sc and} and {\sc xor} function.  From \cite{adamatzky2012architectures}.
 (k) Half-adder made from uniform BZ vesicles. From \cite{adamatzky2012architectures}.
}
\label{fusiongate}
\end{figure}

The collision-based gate shown in Fig.~\ref{CollisionGate} can be also implemented in a geometrically constrained medium, where wave-fragments interact at the junctions of the excitable channels (Fig.\ref{fusiongate})~\cite{adamatzky2015binary}. In an excitable mode stimulation of a single (Fig.~\ref{excitable10}, input $x$) input leads to wave-fronts propagating to all channels. In the sub-excitable mode a wave-fragment from the input channel propagates ballistically to an output channel (Fig.~\ref{subexcitable10}) but when both input channels are excited the wave-fragments fuse and propagate into the central vertical channel (Fig.~\ref{subexcitable11}). The fusion gates can be cascaded into a one-bit half-adder (Fig.~\ref{bzhalfadder}--\ref{bzhalfadder11}), and further to a many-bit full adder~\cite{adamatzky2015binary} and Fredkin and Toffoli logically reversible gates~\cite{adamatzky2017fredkin}.

A fine-grained compartmentalisaiton of BZ solutions can be achieved by preparing emulsions or liquid droplets of BZ solution in an oil and liposomes~\cite{vanag1996ph,vanag2004waves,delgado2011coupled,vanag2011excitatory,costello2012initiation,tomasi2014chemical,torbensen2017chemical}. Then each of the droplets can be seen as computing element interacting with its neighbours via reagents diffusing in oil.  Information transmission, as presented by excitation wave fronts, between BZ droplets~\cite{tomasi2014chemical,gruenert2015understanding,torbensen2017chemical} has been demonstrated in laboratory experiments.  Experimental laboratory computing devices made from arrays of BZ droplets or vesicles are the memory devices in BZ-oil emulsion~\cite{kaminaga2006reaction} and the {\sc nor} gate made from BZ droplets with inhibitory coupling~\cite{wang2016configurable}. Recently, we demonstrated prototypes of BZ liquid marbles, a BZ solution coated with a hydrophobic powder~\cite{fullarton2018belousov}. There are evidences of oxidation wave-fronts transmission between BZ liquid marbles, possibly via gaseous phase, however more experimental evidences are required to establish design of potential computing architectures. 

As demonstrated in computer experiments, by arranging a configuration of BZ droplets and by tuning sizes of pores between BZ vesicles we can implement Boolean gates, including collision-based polymorphic gates~\cite{adamatzky2011polymorphic} and binary arithmetic circuits~\cite{adamatzky2012architectures,holley2011logical,holley2011computational,adamatzky2014logical}. In the example shown in Fig.~\ref{BZDropletAdder} a one-bit half-adder is implemented in a simulated ensemble of two-dimensional BZ vesicles. The  outputs from the {\sc xor} operation are recombined with an {\sc or} operation with additional discs in the top right. The circuit employs three methods of signal modulation: connection angle, disc size and aperture efﬁcacy. Vesicles sizes and diameters of pores between vesicles are selected intentionally to perform the desired operations. The half-adder implemented in uniformly size vesicles having the same diameters of pores is shown in Fig.~\ref{BZDropletAdder_v2A}.

Recently, Gorecki et al.~\cite{gorecki2014information} demonstrated  how to make a frequency of oscillation based {\sc or} gate and one-bit memory in an arrays of couple oscillatory BZ droplets and to evolve a classifier of BZ droplets~\cite{gizynski2017evolutionary}.

BZ reaction is not the only chemical medium for unconventional computing. In mid-1990s Adamatzky and Tolmachiov developed prototypes of reaction-diffusion computing devices capable of approximation of Voronoi diagram of a planar point set~\cite{tolmachiev1996chemical} and calculation of a skeleton of a planar shape~\cite{adamatzky1997chemical}.  Several experimental laboratory prototypes of precipitating processors have been built for implementation of {\sc xor} gate~\cite{adamatzky2002experimental}, computational of a skeleton of a planar shape~\cite{adamatzky2002experimental} and Voronoi diagram~\cite{de2004formation}.

\section{Discussion}
 
 In the past, fluids played a key yet short living role in designs of computing devices. After being outnumbered by electrical analog computers and then digital computers the fluid computers found themselves confined to a narrow range of military applications and control units of nuclear power stations. The liquid based computing, sorting and actuating devices suddenly resurfaced in at a micro-scale, in microfluidics, and in a digital representation, liquid droplets and liquid marbles. In this brief, and possibly subjective, review we touched all  principle developments in liquid computing but liquid electronics.  Breadth and width of the liquid electronics `eco-system' requires a dedicated review paper. Example `species' of the liquid electronics devices are  liquid-solid phase reversible change of conductivity controlled by temperature~\cite{zheng2011reversible}, field controlled electrical switch~\cite{wissman2017field},  liquid field effect transistor~\cite{kim2008iv}, optical liquid droplet switch~\cite{ren2010optical}. There is a also a huge, and largely unexplored, potential in combining mechanical properties of the liquid jets, droplets and marbles with electrical and optical  properties of their solvents and solutes.

%%%%%%%%%% Insert bibliography here %%%%%%%%%%%%%%

%\begin{thebibliography}{9}

%\end{thebibliography}

\bibliographystyle{plain}
%\bibliographystyle{abbrv}
%\twocolumn
\bibliography{biblio}

\end{document}